\documentclass[prd, twocolumn, lengthcheck, superscriptaddress, showpacs, letterpaper, nofootinbib]{revtex4-1}

\usepackage{latexsym}
\usepackage{graphicx}

\usepackage{color}

\usepackage{amsmath}
\usepackage{amssymb}
\usepackage{hyperref}
\usepackage{bm}
\usepackage{acronym}
\usepackage{url}
\usepackage[normalem]{ulem}   
\usepackage{xspace}
\usepackage{subcaption}

\usepackage{amsopn} 

\newcommand{\gws}{\tilde{h}}
\newcommand{\scf}{\ensuremath{\mathcal{C}}}
\newcommand{\scft}{\scf\xspace}

\newcommand{\btheta}{\bm{\theta}}

\newcommand{\dcc}{LIGO-P1500198}

\newcommand\pasp{PASP}

\begin{document}

\title{Astrophysical calibration of gravitational-wave detectors}

\author{M.~Pitkin}
\email{matthew.pitkin@glasgow.ac.uk}
\affiliation{SUPA, School of Physics and Astronomy, University of
  Glasgow, Glasgow G12 8QQ, United Kingdom}
\author{C.~Messenger}
\email{christopher.messenger@glasgow.ac.uk}
\affiliation{SUPA, School of Physics and Astronomy, University of
  Glasgow, Glasgow G12 8QQ, United Kingdom}
\author{L.~Wright}
\altaffiliation{Now at: School of Mathematics and Physics, Queen's University Belfast, Belfast BT7 
1NN, United Kingdom}
\affiliation{SUPA, School of Physics and Astronomy, University of
  Glasgow, Glasgow G12 8QQ, United Kingdom}
  
\date{\today}

\begin{abstract}
  We investigate a method to assess the validity of gravitational-wave detector calibration 
through the use of gamma-ray bursts as standard
  sirens. Such signals, as measured via gravitational-wave
  observations, provide an estimated luminosity distance that is
  subject to uncertainties in the calibration of the data.  If a host
  galaxy is identified for a given source then its redshift can be
  combined with current knowledge of the cosmological
  parameters yielding the true luminosity distance.  This will then allow
  a direct comparison with the estimated value and can validate the accuracy of the original 
calibration. We use simulations of individual detectable gravitational-wave signals from binary 
neutron star (BNS) or neutron star-black hole systems, which we assume to be found in 
coincidence with short gamma-ray bursts, to estimate any discrepancy in the overall scaling of the 
calibration for detectors in the Advanced LIGO and Advanced Virgo network. We find that the 
amplitude scaling of the calibration for the LIGO instruments could on average be confirmed to 
within $\sim 10\%$ for a BNS source within 100\,Mpc. This result is largely independent of the 
current detector calibration method and gives an uncertainty that is competitive with that expected 
in the current calibration procedure. Confirmation of the calibration accuracy to within 
$\sim 20\%$ can be found with BNS sources out to $\sim 500$\,Mpc.
\end{abstract}

\preprint{\dcc}

\maketitle

\acrodef{GW}[GW]{gravitational-wave}
\acrodef{NS}[NS]{neutron star}
\acrodef{BNS}[BNS]{binary neutron star}
\acrodef{MWEG}[MWEG]{Milky Way equivalent galaxies}
\acrodef{BBH}[BBH]{binary black hole}
\acrodef{BNS}[BNS]{binary neutron star}
\acrodef{NSBH}[NSBH]{neutron star-black hole}
\acrodef{SNR}[SNR]{signal-to-noise ratio}
\acrodef{GRB}[GRB]{gamma-ray burst}
\acrodef{sGRB}[sGRB]{short gamma-ray burst}
\acrodef{CBC}[CBC]{compact binary coalescence}
\acrodef{EM}[EM]{electromagnetic}
\acrodef{aLIGO}[aLIGO]{Advanced LIGO}
\acrodef{AdV}[AdV]{Advanced Virgo}
\acrodef{PSD}[PSD]{power spectral density}
\acrodef{PDF}[PDF]{probability distribution function}
\acrodef{MCMC}[MCMC]{Markov chain Monte Carlo}
\acrodef{ET}[ET]{Einstein Telescope}
\acrodef{XRT}[XRT]{x-ray telescope}

\section{Introduction\label{sec:intro}}

It is expected that the advanced generation of interferometric \ac{GW}
detectors will detect waves emitted from $O(10s)$ of \acp{CBC} per year
\cite{2010CQGra..27q3001A}. One such class of these cataclysmic
events, the inspiral and merger of \ac{BNS} systems will be detected out to a
maximum range (horizon distance) of $\approx 450$ Mpc. Assuming the
current best estimates for the cosmological parameters, this is equivalent to a
redshift $z\approx 0.1$. As noted by~\citet{1986Natur.323..310S}, \ac{CBC}
systems can be used as cosmological distance markers, otherwise known as
``standard sirens'' (analogous to \ac{EM} standard candles). The nature of
these sources will allow us to estimate, among other parameters, their
luminosity distance without the need for calibration against the cosmic
distance ladder.

It is considered likely that the merger of \ac{BNS} and/or \ac{NSBH} systems,
in addition to emitting detectable \acp{GW}, is also the mechanism for
producing \acp{sGRB} \cite{1992ApJ...395L..83N}. In this scenario these events
produce tightly beamed \ac{EM} emission parallel to the orbital angular
momentum vector of the system. Additional evidence for the coincidence of
\acp{sGRB} with \ac{GW} events is the estimated astrophysical rates of both
phenomena. Observations of \acp{sGRB} give an inferred rate of \ac{sGRB}
producing \ac{CBC} mergers of $8\times 10^{-9}$--$1.1\times
10^{-6}$\,Mpc$^{-3}$\,yr$^{-1}$~\cite{2012MNRAS.425.2668C} after accounting for
beaming effects. This can be compared to independently obtained estimates of
\ac{BNS} merger rates of $10^{-8}$--$10^{-5}$\,Mpc$^{-3}$\,yr$^{-1}$
\cite{2010CQGra..27q3001A}, or \ac{NSBH} merger rates of
$6\,\times\,10^{-10}$--$1\,\times\,10^{-6}$\,Mpc$^{-3}$\,yr$^{-1}$
\cite{2010CQGra..27q3001A}\footnote{However, the recent work of
\cite{2015MNRAS.448..928K}, which studied known galactic \ac{BNS} systems in
more detail, suggests a rate approximately five times lower.}.
  
If a single \ac{GW} event is observed in coincidence with a \ac{sGRB} it may be
possible to identify the host galaxy of the source.  With this information a
spectroscopic redshift can be very accurately obtained (see values quoted in
e.g.\ \cite{0004-637X-664-2-1000} and references therein). With knowledge of
the redshift, using current best estimates for the Hubble constant and other
cosmological parameters, the true luminosity distance can be estimated to $\sim
1\%$ accuracy~\cite{2015arXiv150201589P}.  The \ac{GW} measurement
acts as a standard siren also giving us a direct measurement of the luminosity
distance to the source. The accuracy of such a measurement depends on a number
of factors including the accuracy with which the \ac{GW} detector has been
calibrated.  Hence, by comparison with the distance estimate from the \ac{sGRB}
we can recalibrate (or validate) the existing experimentally obtained
calibration. 

In this paper we investigate the feasibility of this approach for single coincident 
\ac{GW}--\ac{sGRB} events and establish the validation power of such a calibration technique.  In 
Sec.~\ref{sec:calibration} we briefly summarize the existing experimental technique for \ac{GW} 
detector calibration and its expected accuracy.  We then review the concept of \ac{GW} standard 
sirens and their proposed coincident \ac{EM} signatures, the \acp{sGRB} in
Secs.~\ref{sec:sirens}--\ref{sec:GRB}. In Secs.~\ref{sec:analysis}--\ref{sec:results} we 
discuss our analysis method and the results, and in Sec.~\ref{sec:discussion} we 
conclude.    

\section{Gravitational-wave detector calibration\label{sec:calibration}}

The technique used throughout \ac{GW} research in calibrating the detectors is
carried out through a complicated system of physical manipulations of the
Fabry-P\'{e}rot Michelson interferometer, where an elaborate feedback
system is used to sustain a defined measurement in arm length difference
between the moving mirrors. A comprehensive description of the calibration
procedure (in particular for the LIGO detectors during their fifth science run)
can be found in \cite{2010NIMPA.624..223A}, and a brief description is given in
\cite{Vitale:2012} and references therein. For \ac{aLIGO} and \ac{AdV}, at the
frequency range used in this analysis (20--400 Hz), the error in the
strain amplitude calibration is expected to be roughly $10\%$
\cite{Vitale:2012}. This is a benchmark set for the error estimation using the
proposed method in this project.

The \ac{GW} strain is measured through the differential arm length, $\Delta L$,
changes of the interferometer via $h(f,t) = \Delta L(f,t) / L$, where $L$ is
the full arm length. Calibration is required to relate the actual measured
interferometer error signal output $e(f, t)$ to $\Delta L$.  This relation is
known as the length response function, $R(f)$, defined such that
\begin{equation}
\Delta L(f,t) = R(f) e(f, t),
\end{equation}
where we assume $R$ varies only slowly in time (in comparison to transient
signal time scales). Calculation of $R$ requires the measurement of various
functions within a control feedback loop (see \cite{2010NIMPA.624..223A}),
each of which are subject to measurement uncertainties. In this study
we assume an estimate of $R$ is available (although in theory we could
take on the role of estimating $R$ itself), but that it differs from the truth
through some unknown scale factor, $\scf$, so that at a particular time we have
\begin{equation}\label{eq:scalefactor}
\scf h(f) = h_{\rm m}(f) = \frac{R(f) e(f)}{L},
\end{equation}
where $h(f)$ is the true strain and $h_{\rm m}(f)$ is the measured strain
(i.e.\ the measured length response function $R(f)$ is related to the
true length response function $R_{\rm true}(f)$ via $R(f) = \scf R_{\rm
true}(f)$). With this definition it means that if $\scf > 1$ then a signal
would appear to have a larger amplitude (i.e.\ for a given system,
would seem closer) than in reality\footnote{However, the \ac{SNR} would be
the correct value as the signal and the noise will both contain the same scale
factor.}, whereas if $\scf < 1$ it would appear to have a lower amplitude than
reality.  In this analysis we simplify the situation by assuming that
$\scf$ is a constant with respect to frequency (and is real, so has no phase
component), but in future studies that assumption could be dropped and $\scf$
could take some functional form, or piecewise fit, with respect to $f$ [e.g.\
in a similar way to the method of \cite{2013PhRvD..88h4044L} used for fitting
differences in \ac{PSD} estimates].

\section{Binary neutron star standard sirens\label{sec:sirens}}

The idea of \ac{GW} standard sirens is directly analogous to the concept of
standard candles in \ac{EM} astronomy and was first proposed
in~\cite{1986Natur.323..310S}. Unlike the primary \ac{EM} standard candle
event, type Ia supernovae, the measured luminosity of a \ac{CBC} in \acp{GW} is
not only a function of distance. It also depends upon the chirp mass (a
function of the component masses) and the binary orientation with respect to
the global network of \ac{GW} detectors. However, measurement of the
phase evolution of such an event allows accurate determination of the chirp
mass\footnote{In reality it is the redshifted chirp mass that is measured.}.
Timing information from the different signal arrival times at each detector in
the network allows some level of sky position
determination~\cite{2014ApJ...795..105S}.  Finally, amplitude variation between
differently oriented interferometers allows some (weaker) level of
determination of the binary system orientation.

Standard sirens are clearly a very powerful tool for \ac{GW} cosmology since
they give us a direct measure of the absolute luminosity distance to sources.
This is distinct from type Ia supernovae standard candles that only provide
relative luminosity distance measures and require calibration via other methods
as part of the cosmological distance ladder. The use of \ac{GW} events with
\ac{sGRB} counterparts for use as a cosmological tool was investigated
in~\cite{2010CQGra..27u5006S}--\cite{2011PhRvD..83b3005Z} with respect to
the third generation \ac{GW} interferometer, the
Einstein Telescope~\cite{2010CQGra..27h4007P} and in ~\cite{2012PhRvD..85b3535T}
for the Advanced detector network. In this case, the redshift information
obtained from the \ac{sGRB} host galaxy provides a complementary measurement
allowing each event to inform the distance-redshift relationship and hence
obtain cosmological parameter estimates. Other methods for cosmological
inference (see~\cite{1996PhRvD..53.2878F,2012PhRvD..85b3535T,
2012PhRvD..86b3502T,Messenger:2011ux,2013arXiv1312.1862M}) have been proposed
for \ac{BNS} systems that do not rely on any redshift measurements. These
instead use either the distribution of measured \acp{SNR} and assumed
form of \ac{NS} mass and spatial distributions or exploit the
features of the tidal and postmerger hypermassive \ac{NS} stages of the
\ac{GW} waveform. 

To put the existing work on \ac{GW} cosmology in context with the study
described in this manuscript we are essentially turning the standard
cosmological problem upside down. We assume that the existing \ac{EM}
cosmological parameter estimation [specifically for the Hubble constant, which
has uncertainties of $O(1\%)$~\cite{2015arXiv150201589P}] is
accurately determined and therefore a known quantity. We then assume that the
unknown in our analysis is the absolute amplitude calibration of the
interferometers in the global \ac{GW} detector network.    

\section{GRB counterparts\label{sec:GRB}}

It is believed that \acp{sGRB} (those gamma-ray bursts with duration ${<}2$ sec) are
emitted during the merger of \ac{BNS} or \ac{NSBH} systems
~\cite{1989Natur.340..126E,1992ApJ...395L..83N}. The emission from these events
is highly beamed along the binary rotation axis and hence only potentially
observable for a fraction of the mergers. If the event exhibits an optical
afterglow then the host galaxy can be identified, from which a redshift can
be obtained (e.g.~\cite{2005Natur.437..851G}).  The fraction of events with
associated redshifts is
${\sim}1/3$~\footnote{http://swift.gsfc.nasa.gov/archive/grb\_table/}.
The range to which the \emph{Swift} x-ray telescope has detected \acp{sGRB} is
$z{\sim}2$ and the nearest event with associated redshift is at
$z{\sim}0.1$~\cite{2015GCN..17278...1C}, which is approximately equal to the
horizon range of the advanced \ac{GW} detector network for \ac{BNS} systems. 

The predicted detection rates of \ac{BNS} events (irrespective of an \ac{sGRB}
counterpart) are derived from three main sources: population
synthesis models, extrapolations based on known galactic \ac{BNS} systems, and
observations of \ac{sGRB} events. Various studies using the first two
methods have been compiled into an overall detection rate of $0.2$--$200$
\ac{BNS} events per
year~\cite{2013arXiv1304.0670L,2010CQGra..27q3001A}\footnote{We note that this
rate is that given in Table I of \cite{2013arXiv1304.0670L} for an \ac{aLIGO}
and \ac{AdV} network consisting to two \ac{aLIGO} detectors operating at design
sensitivity. However, a higher rate of $0.4$--$400$ is given in
\cite{2010CQGra..27q3001A}, which may differ due to the use of a third
\ac{aLIGO} detector, but the discrepancy may also have contributions from
slightly different threshold definitions or design sensitivity curves used. We
also note the study mentioned in Sec.~\ref{sec:intro} that potentially lowers
the expected rate by a factor of five.}. A similar event rate of
$1$--$180$ was obtained in~\cite{2012MNRAS.425.2668C}, where the rate of
observed \acp{sGRB} was converted to the rate of \ac{BNS} events by assuming a
beaming angle of $15^{\circ}$. The beaming angle is the primary factor in
determining the fraction of detected \ac{GW} \ac{BNS} signals with an \ac{EM}
\ac{sGRB} counterpart and is unfortunately highly uncertain.  Various authors
have provided estimates on the rate of joint detections to be in the range
$0.02$--$7$ per
year~\cite{2013PhRvL.111r1101C,Kelley:2012fl,2014arXiv1405.2254S,2015MNRAS.448.3026W}.  

The most likely scenario in which a coincident \ac{GW}--\ac{sGRB} event would
be identified is through the targeted follow-up of an observed \ac{GW} or by
\emph{post facto} matching of \ac{GW} trigger lists with known (or subsequently found
\cite{2015ApJS..217....8B}) \ac{sGRB} events.  The likelihood of being able to
follow-up \ac{GW} events using gamma-ray telescopes with low enough latency to
catch a \ac{sGRB} is low.  Compounding this issue is the relatively large
\ac{GW} sky error box giving a field of view for the \ac{EM} observatories to
search spanning $\sim 100$s of square degrees (e.g.\
\cite{2011CQGra..28j5021F,2013arXiv1304.0670L,2014ApJ...795..105S}).
For the scenario of a \ac{GW}-followup of an observed \ac{sGRB}, the
merger time for \ac{BNS}/\ac{NSBH} systems will be estimated from the \ac{sGRB}
to within a few seconds (see e.g.\ the discussion in Sec.\ 2.2 of
\cite{2012ApJ...760...12A}). This makes the follow-up search less
computationally expensive since it is performed over a
smaller range of data using potentially fewer numbers of waveform templates.
This computational saving enables the use of a more computationally expensive
multi-detector coherent scheme rather than the cheaper coincidence methods used
in the untargeted searches. 

A fortunate consequence of a joint \ac{GW}--\ac{sGRB} observation will be the
fact that in order for such an event to be observed, the \ac{BNS}/\ac{NSBH}
system must have had its orbital angular momentum vector pointing towards (or
away) from the detector.  The actual inclination angle of the system, defined
as the angle between the orbital angular momentum vector and the line of sight,
must be $<$ half of the \ac{sGRB} beaming angle.  This prerequisite property
limits us to systems that are approximately ``face-on'' and therefore
effectively optimally oriented. At given distances this makes us biased towards higher
\ac{SNR} signals but it is more meaningful to think of this as an increase in
the sensitivity range of the detectors. However, as discussed earlier, the
property of beaming severely impacts the probable rate of such joint
observations.

\section{Analysis}\label{sec:analysis}

The main aim of this work is to assess how well the calibration scale factor
defined in Eq.~\ref{eq:scalefactor} can be estimated from a single observed
\ac{GW} associated with a particular \ac{sGRB}. As we {\it a priori} have
little knowledge of the likely location and distance of such an event we have
performed simulations of multiple events to see the distribution in the accuracy of the calibration 
scale factor recovery.

For all our signals we use the TaylorF2 waveform approximation (see e.g.~\cite{2009PhRvD..80h4043B}
and references therein) with a 3.5 post-Newtonian expansion in phase for
modeling our signal (in both simulations and signal recovery).  For \ac{BNS}
systems we use nonspinning waveforms under the assumption that the component spins will be small 
and have a negligible effect on the analysis\footnote{In future studies we could create our 
simulations including small spins and confirm that this does not have a significant effect on our 
results if recovery is with nonspinning templates.} (see discussion in \cite{2014ApJ...795..105S} 
and studies in \cite{2012PhRvD..86h4017B}). For \ac{NSBH} systems we use a spinning waveform, but in 
which only the black hole has non-negligible spin and its rotational angular momentum is aligned 
with the system's orbital angular momentum.

For a nonspinning system the general form of the frequency-domain polarization
amplitudes is
\begin{align}\label{eq:signal} 
\gws_{+}(f) &\propto
\left(1+\cos^{2}\iota\right)D_{L}^{-1} \mathcal{M}^{5/6}f^{-7/6}e^{-i\Psi(f,
\mathcal{M}, t_\mathrm{c}, \phi_{\mathrm{c}})}, \nonumber \\
\gws_{\times}(f) &\propto
\cos{\iota}D_{L}^{-1}\mathcal{M}^{5/6} f^{-7/6}e^{-i\Psi(f, \mathcal{M},
t_\mathrm{c}, \phi_{\mathrm{c}})} 
\end{align}
where the chirp mass $\mathcal{M}$ is defined as $\mathcal{M}=Mq^{3/5}$,
with the symmetric mass ratio $q=m_{1}m_{2}/M^{2}$, total mass
$M=m_{1}+m_{2}$ (where $m_{1}$ and $m_{2}$ are the component masses), and $\iota$ is
the binary inclination angle.  The Fourier transform of the \ac{GW} strain measured
at the $k^{\rm{th}}$ detector is then given by
\begin{equation} \label{eq:gravsig} 
\gws^{k}(f) = F_{+}^{k}(\alpha,
\delta, \psi)\gws_{+}(f) + F_{\times}^k(\alpha, \delta,
\psi)\gws_{\times}(f)
\end{equation}
where $F^{k}_{+}$ and $F^{k}_{\times}$ are the antenna response functions which
are dependent upon the polarization angle $\psi$ and the sky position of the
source with right ascension $\alpha$ and declination $\delta$.

For this analysis we consider the advanced generation \ac{aLIGO} and
\ac{AdV} detectors operating at their design sensitivity, with noise power
spectral densities taken from \cite{2013arXiv1304.0670L}. The detector network
we consider consists of the two \ac{aLIGO} detectors (H1 and L1) and 
the \ac{AdV} detector (V1). 

We consider that the measured data in any detector are defined as
\begin{equation}
  \tilde{d}_{k}(f)=\scf_{k}(\tilde{n}_{k}(f)+\gws_{k}(f,\btheta))
\end{equation}
where $\tilde{n}(f)$ is the Fourier transform of the true (in this case
Gaussian) strain noise, $\btheta$ is the set of waveform parameters and
$\scf_k$ is the calibration scale factor for the $k^{\rm th}$ detector, which are the
parameters we are interested in estimating.

\subsection{Method}

To assess the ability to estimate the unknown calibration scale factors we
calculate their marginal posterior \acp{PDF}.  We use Bayes' theorem for which
we need to define a likelihood function and prior \acp{PDF} on the signal
parameters and the calibration scale factors. In this analysis we use the
\ac{MCMC} code {\tt emcee}~\cite{2013PASP..125..306F} to sample
the posterior distribution. We use a Gaussian likelihood function given by
\begin{widetext}
\begin{equation}
p(\bm{d} | \btheta, \bm{\scf}, I) \propto \exp{\left[4\Delta f \sum_{k=1}^{N_{\rm 
det}}
  \sum_{i=i_{\rm low}}^{i_{\rm high}}\frac{\Re{\left\{ \tilde{d}^{*}_{k,i} 
\scf_{k}\gws_{k,i}(\btheta) - \frac{1}{2}\left(\scf_k^2 \gws^*_{k,i}(\btheta)\gws_{k,i}(\btheta) + 
\tilde{d}^*_{k,i}\tilde{d}_{k,i}\right) \right\}}}{S_{k,i}}\right]}
\end{equation}
\end{widetext}
where $\bm{d}$ is an array containing the data for all detectors, $\Delta f$ is
the frequency resolution, $S_k$ is the $k^{\rm th}$ detector's estimated
one-sided noise \ac{PSD} (and as such will contain the effect of the
calibration scale factor), and the $i$ indices increment over frequency with
$i_{\rm low}$ and $i_{\rm high}$ corresponding to the lower and upper range in
frequencies used for our analysis, which were 20 to 400\,Hz respectively.
This frequency range was chosen as the vast majority of the \ac{SNR} for the
inspiralling signals we use can be found within it (for a typical \ac{BNS}
system of two $1.4\,{\rm M_{\odot}}$ stars, or a \ac{NSBH} system with component masses of
$5\,{\rm M_{\odot}}$ and $1.4\,{\rm M_{\odot}}$, only $\sim 3-4\%$ more \ac{SNR} is gained
by using a 10 to 1500\,Hz frequency range), so going to lower
or higher frequencies provides very little additional information while increasing
the computational cost\footnote{In a real analysis, rather than this case study, the additional
computational cost of using a larger frequency range may be worthwhile, but in this case we had
to run many simulations and the computational cost was a limiting factor.}. The source
parameters have been subsumed into a vector $\btheta$ and the calibration scale
factors are within the vector $\bm{\scf}$.

\subsubsection{Prior ranges}\label{sec:priors}

The primary reason why coincident observations with a \ac{sGRB} can be used as
a check of detector calibration is that the \ac{sGRB} allows us to constrain
the priors on various parameters that would normally be highly correlated with
the signal amplitude. The most important of these is that the \ac{sGRB} can
provide a very tight constraint on the source distance. In this analysis we therefore
assume that the source distance is known (i.e.\ has a $\delta$-function
prior) whereas in reality the error on this distance will be dependent
upon the uncertainties in both the redshift measurement and the Hubble
constant. For spectroscopic redshifts this uncertainty will be dominated by the
Hubble constant and will be of $O(1\%)$.

Other information relevant for a coincident \ac{sGRB} is that the system is
likely to be relatively close to face on (and therefore nearly circularly
polarized), which allows us to place prior constraints on the system
inclination angle. Beam opening angles for \acp{sGRB} are relatively poorly
constrained as they are based on only a few jet-break detections. A comparative
study of estimates and lower limits of opening angles for 13 \acp{sGRB} given
in~\cite{2014ApJ...780..118F} provides a median opening angle value of $\sim
10^{\circ}$. We use this median opening angle as a rather conservative
guide to form a Gaussian prior distribution on the system inclination angle,
$\iota$, by converting it into an equivalent standard deviation for a
half-normal distribution. We therefore place a Gaussian prior on $\iota$ with
zero mean and a standard deviation $\sigma=14.8^{\circ}$. This, and the choice
of polarization angle prior range given below, explicitly fixes the handedness
of the system's rotation (or alternatively whether it is face on or back on).
If we were to allow the inclination to have two modes (one for face on and one
for back on), or alternatively to double the range of the polarization angle,
it would have no effect on our ability to estimate the signal amplitude as the
two modes should be identical with regards to the amplitude parameter
probability distributions. However, we are consistent between the range of
simulated signals discussed in Sec.~\ref{sec:simulations} and the initial
ranges stated.

The prior on the system component masses is dependent on whether we are
considering the source being a \ac{BNS} system or a \ac{NSBH}. For the former
case the prior we use is a Gaussian distribution for both components with means
of $1.35\,\textrm{M}_{\odot}$ and standard deviations of
$0.13\,\textrm{M}_{\odot}$, which is roughly consistent with the double neutron star system
masses found in \cite{2012ApJ...757...55O} albeit with a wider distribution. In reality, for all 
the systems we use in our study, the chirp mass, which we see from Eq.~\ref{eq:signal}
plays a role in the overall signal amplitude, is very well constrained by the
system's phase evolution, so our prior could be expanded with minimal effect on
the results. In the case of \ac{NSBH} systems we use the same prior as above
for the neutron star, but for the black hole we use a prior based on the
canonical mass distribution used for the rate results in
\cite{2012PhRvD..85h2002A} with a mean of $5\,\textrm{M}_{\odot}$ and a
standard deviation of $1\,\textrm{M}_{\odot}$. We note that the mass range of
black holes could be quite different from this, but use this range as an
example for this type of system. For the \ac{NSBH} systems we also require a
prior on the black hole spin (for \ac{BNS} systems we have assumed that the
spins are small enough that they will be negligible). We use a uniform prior on
the normalized aligned spin magnitude between $-1$ and $1$.

The final important piece of information that we can make use of from the
\ac{sGRB} observation is the sky position of the source. We assume that this is 
known precisely and has been obtained, in conjunction with the source redshift,
from the identification of the host galaxy. This fixes the time delay between detectors to a known 
value and, along with the polarization angle, defines the antenna response patterns.

For the time of coalescence $t_{\mathrm{c}}$ we use a uniform prior spanning
$\pm0.01$\,seconds around the recorded time of the signal that would be
returned by the \ac{GW} detection pipeline. For the reference phase
$\phi_{\mathrm{c}}$ a uniform prior on the range $(0,\pi]$ is used, and for the
polarization angle we use a uniform prior on the range $(0,\pi/2]$. As the
signals we use are generally close to being circularly polarized (face on) the
phase and polarization angle will be largely degenerate and their exact
combination will have little effect on the result.

Finally, we require a prior on the calibration scale factors for each detector.
We make the assumption that in general the calibration applied to the detector
data will be close to being correct, so we want to use a prior that is peaked at $\scf_k=1$. We
also (na\"{i}vely) assume that the prior \ac{PDF} of the calibration scale factor is either
larger or smaller than unity by an equivalent factor, e.g.\ $p(\scf=10|I)$ is
equivalent to $p(\scf=0.1|I)$. For the scale factor in each detector we 
therefore use the following log-normal distribution as our prior,
\begin{equation}\label{eq:scale_prior}
 p(\scf|I) = \frac{1}{\scf\sigma\sqrt{2\pi}}\exp{\left( -\frac{(\ln{\scf} - \mu)^2}{2\sigma^2} 
\right)},
\end{equation}
where we chose $\sigma = 1.07$, which together with the requirement that the
mode is at unity gives $\mu = 1.15$. With these parameters the prior
probability density at 0.1 and 10 is one tenth of that at 1 making this a fairly
conservative prior designed to have little impact on our results. Note that this
prior does not give a symmetric amount of probability about unity (so our prior
favors \scf\xspace values greater than 1), but given our simulation criterion
described in Sec.~\ref{sec:simulations} we find that the likelihood
generally overwhelms the prior and our na\"{i}ve assumptions have little effect.

\subsection{Simulations}\label{sec:simulations}

To estimate how well the calibration scaling can be constrained for each
detector in an advanced detector network we have performed analyses with simulated
\ac{BNS} and \ac{NSBH} signals. These were simulated
at a range of distances: 50 to 500\,Mpc with 50\,Mpc increments for the
\ac{BNS} signals and 100 to 900\,Mpc with 100\,Mpc increments for the
\ac{NSBH} signals. The simulations used the two \ac{aLIGO} detectors (H1 and
L1) and the \ac{AdV} detector (V1) and were all performed in the frequency
domain and spanned a frequency range from 20 to 400\,Hz, which is a range that contains the 
vast majority of the \ac{SNR} and is therefore all that is required to provide good 
constraints on \scf. We assumed all
detectors were operating at their design sensitivity (as given by Fig.~1 of
\cite{2013arXiv1304.0670L}). As such we added colored Gaussian noise to each
simulation based on the sensitivity curve, but scaled with the associated
calibration scale factor.

Separately for the \ac{BNS} and \ac{NSBH} systems we have performed a large
number [several hundred up to $O(1000)$] of simulations at each distance value.
When proposing injections for each simulation the source parameters were
randomly drawn from the prior distributions described in
Sec.~\ref{sec:priors}, with the sky position being drawn randomly from a
uniform distribution on the sky, the coalescence times held fixed within the
center of the prior window, and with the calibration scaling factors for each
detector drawn from a Gaussian distribution with a mean of one and standard
deviation of $0.125$ (equivalent to a mean fractional offset of 10\%).  Note
that this distribution is broadly consistent with the scale of calibration
uncertainties that is expected via experimental methods, but is not the same as
the prior that we assume on $\scf$ [see Eq.~\ref{eq:scale_prior}] when recovering 
signals\footnote{The reason behind this is that we wanted to use a fairly na\"{i}ve and
conservative prior for signal recovery that had little impact in the results, i.e.\
to produce results domintaed by the likelihood for these parameters. However,
limited studies using $\scf$ values within the much larger range of
$\sim 0.1$ to 10 show distributions with consistent relative widths to
those we find in our main study.}. 

In accepting a proposed simulation as one to be analyzed we introduced a
criterion that the signal be ``detectable'', based on that used in
\cite{2012PhRvD..85h2002A}. We therefore only analyzed simulations that had an
\ac{SNR} of $\geq 5.5$ in at least two detectors. Since we assume that these
signals would be coincident with a \ac{sGRB} we do not include the often used
further constraint that the total coherent network \ac{SNR} is greater than 12.
For simulations close to the \ac{SNR} threshold, the application of the
threshold resulted in the preferential selection of better oriented systems.
Therefore, at larger distances the population of injected
sources was not uniformly distributed over the sky and instead favored locations with better
antenna response.  Similarly, despite already being constrained by the prior to
small inclination angles, the accepted simulations show preference for sources
very close to circularly polarized.

These simulations have allowed us to assess how well on average we would be
able to constrain the calibration scale for a detected \ac{CBC}-\ac{sGRB}
coincidence.

\section{Results}\label{sec:results}

A simple assumption that one could make would be that the uncertainties on the
calibration scale factors (if they are independent of other parameters and have
a roughly Gaussian probability distribution) should be given by $\sim 1/{\rm
SNR}$ for each detector. As we show below this assumption is reasonable,
but small correlations do exist between parameters meaning that it does not
completely hold.

\begin{figure*}
 \begin{center}
  \includegraphics[width=1.0\textwidth]{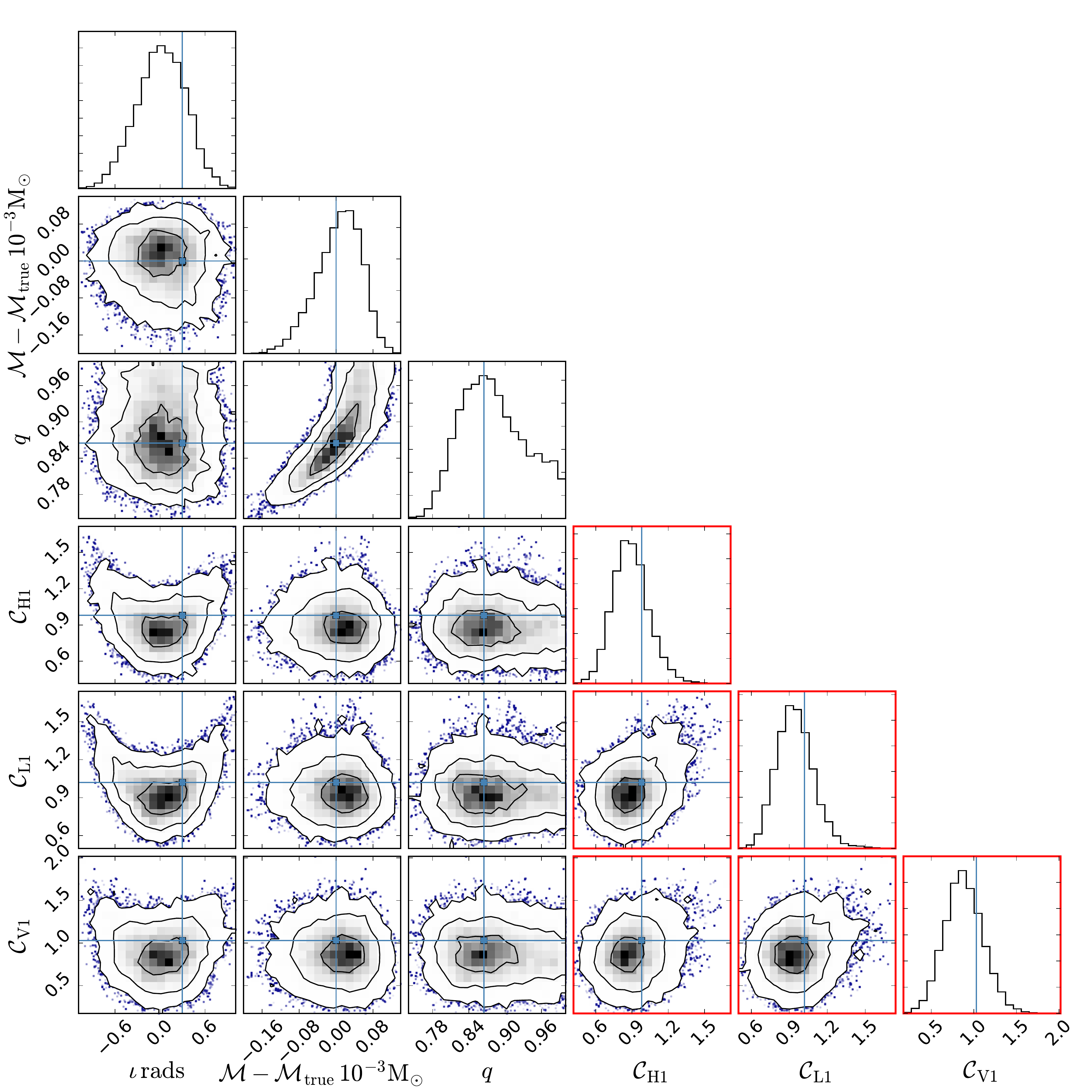}
 \end{center}
 \caption{\label{fig:bnspost} The marginalized posterior probability distributions for some of the unknown
 parameters of a \ac{BNS} system, including calibration scaling factors (surrounded by the 
thicker red 
borders) for the three detectors (H1, L1 and V1). The simulated signal was at a distance of 
250\,Mpc, had \ac{SNR} of 7.9, 9.1 and 5.2 and percentage relative uncertainties in the calibration 
scale factors of 15\%, 14\% and 22\% for each of the detectors respectively.}
\end{figure*}

\begin{figure*}
 \begin{center}
  \includegraphics[width=1.0\textwidth]{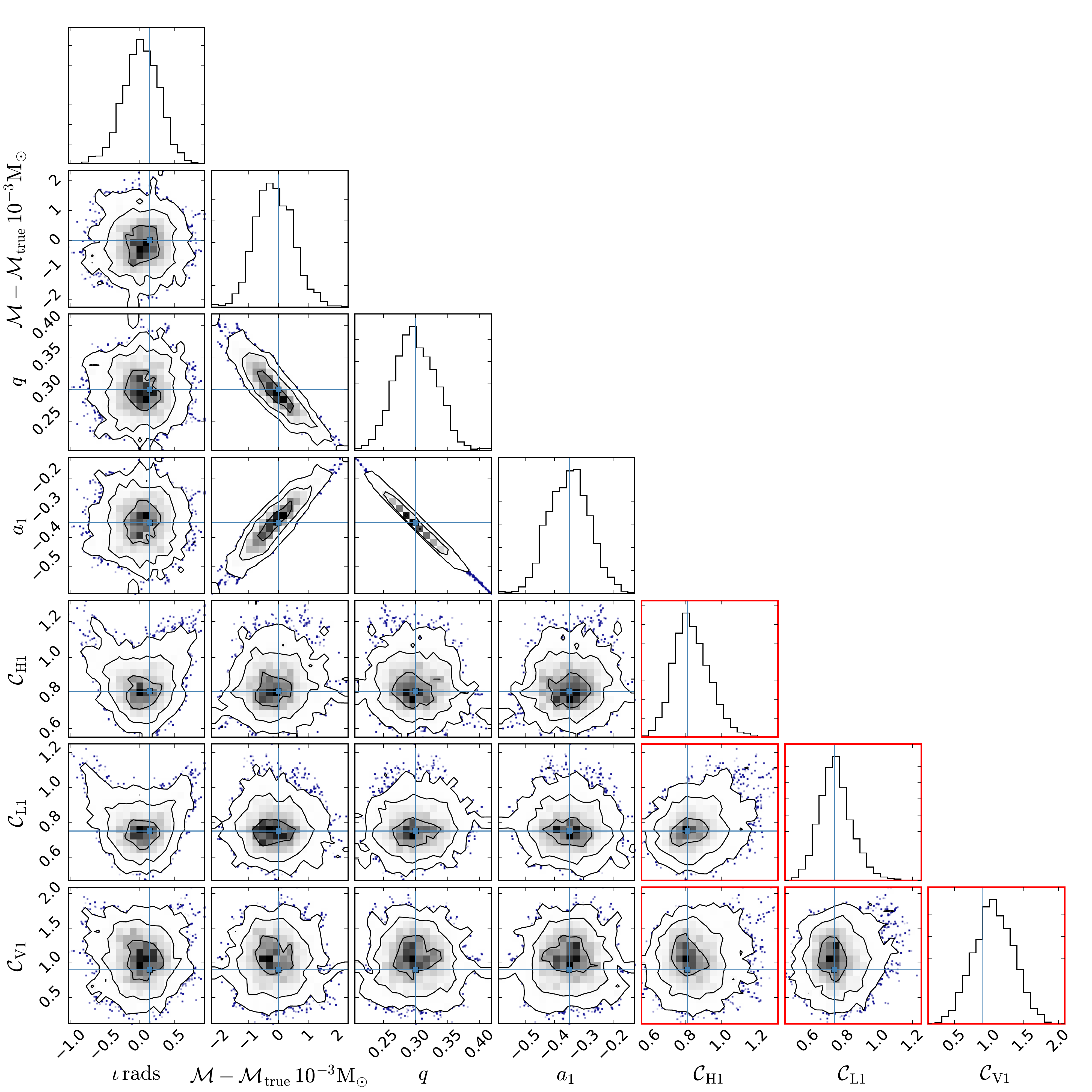}
 \end{center}
 \caption{\label{fig:nsbhpost} The marginalized posterior probability distributions for some of the unknown
 parameters of a \ac{NSBH} system, including calibration scaling factors for the three detectors 
(H1, L1 and V1). The simulated signal was at a distance of 450\,Mpc, had \ac{SNR} of 9.7, 10.3 and 
3.8 and percentage relative uncertainties in the calibration scale factors of 14\%, 13\% and 43\% 
for each of the detectors respectively.}
\end{figure*}

For each of the simulated sources we have used the {\tt emcee} python \ac{MCMC}
package~\cite{2013PASP..125..306F} to perform parameter estimation over the
unknown source parameters using the priors as discussed in
Sec.~\ref{sec:priors}. In each case when calculating the likelihood we used
an estimate of the noise \ac{PSD} based on the advanced detector design
sensitivities (using those given in~\cite{2013arXiv1304.0670L}), but calculated by averaging 
64 separate noisy realizations of the \ac{PSD} and scaled
with the same calibration scale factor as applied to the injection and noise\footnote{We do not account for there 
being a potential difference
between the estimated \ac{PSD} and the actual \ac{PSD} of the analysed section
of data as described in e.g.\ \cite{2013PhRvD..88h4044L}. This difference would
be very highly correlated with the calibration scale factor, so in reality our
estimate of the scale factor would be a combination of the calibration offset
and the difference in the \ac{PSD}. As such our results on real data would be
an upper limit on the calibration scale factor.}. This has provided posterior probability
distributions\footnote{As a proxy to check for convergence of the MCMC chains
we check that the calibration scale factor posterior histograms do not contain many disjoint modes.}
on the calibration scale factors for each detector.  Examples of
the marginalized posterior probabilities for a \ac{BNS} system and a \ac{NSBH}
system observed with the three detector network are shown in
Figs.~\ref{fig:bnspost} and \ref{fig:nsbhpost} respectively.

\begin{figure*}
 \begin{center}
  \includegraphics[width=1.0\textwidth]{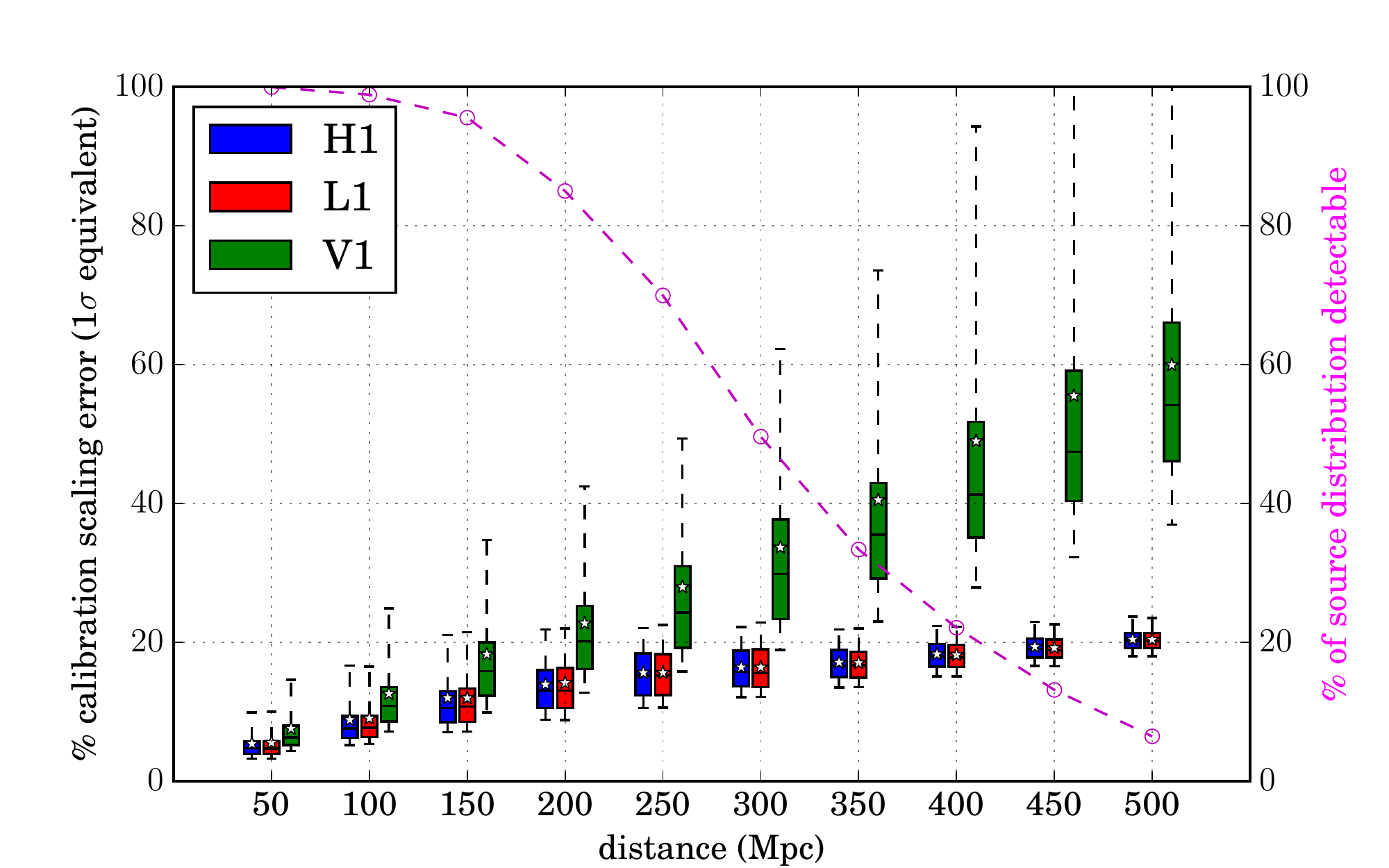}
 \end{center}
 \caption{\label{fig:bnsresults} Distributions of the percentage accuracy at which the calibration 
scale factors can be determined for a three detector network using
\ac{BNS} systems (provided a coincident GRB is observed and can yield a
distance estimate). The boxplots span the lower to upper quartile range of the
distributions, with the median value shown as a horizontal line within the box
and the mean shown as a white star. The dashed magenta line shows the
percentage of sources drawn from the prior distribution that would be detectable
at each distance value.} 
\end{figure*}

\begin{figure*}
 \begin{center}
  \includegraphics[width=1.0\textwidth]{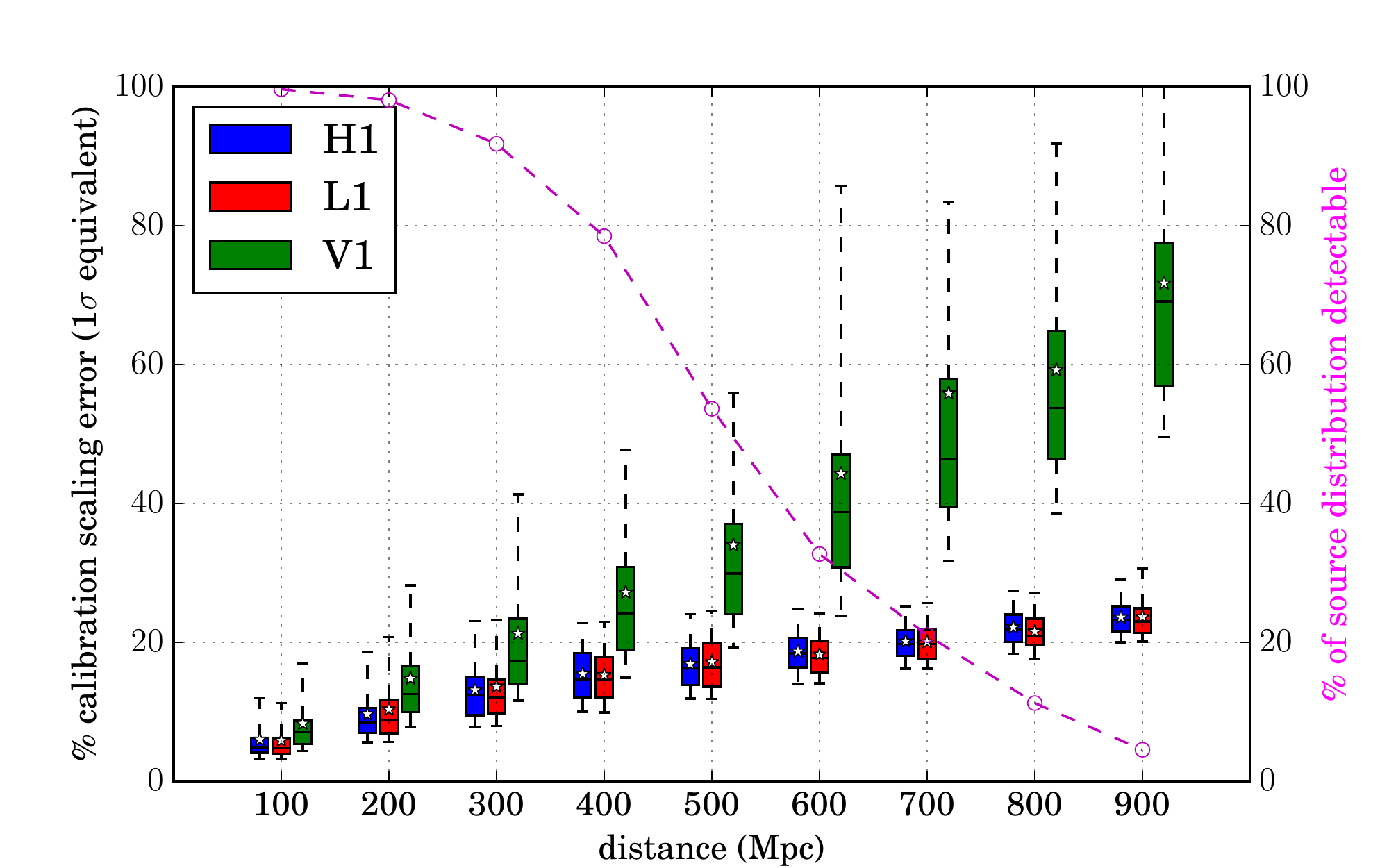}
 \end{center}
 \caption{\label{fig:nsbhresults} Distributions of the percentage accuracy at which the calibration 
scale factors can be determined for a three detector network if using \ac{NSBH} systems 
(provided a coincident GRB is observed and can yield a distance estimate). The plot contents 
are the same as in Fig.~\ref{fig:bnsresults}.}
\end{figure*}

From these posterior distributions we have calculated the minimal 68\% credible
region for the calibration scale factors for each detector (if these were
Gaussian distributions this is equivalent to the region either side of the mean
bounded by the $1\sigma$ intervals). For all the signals at each distance
increment we have produced the distribution of the fractional half widths
(i.e.\ $1\sigma$) of these scale factors' confidence intervals compared to the
true value. These are shown as boxplots in Figs.~\ref{fig:bnsresults} and
\ref{fig:nsbhresults} for the \ac{BNS} and \ac{NSBH} systems respectively. The
boxes show the extent from the lower to upper quartile of the values, while
the whiskers extend from the fifth to 95$^{\rm th}$ percentile. The
black line within each box gives the median value and the star gives the mean
value. Also shown on each plot as the dashed magenta line is the percentage of
signals drawn from the prior distribution that fulfils the detection criterion.

In Fig.~\ref{fig:bnsresults} we see that on average the scale factor can be
recovered to equivalent precision in both \ac{aLIGO} detectors, as would be
expected, with uncertainties generally within the 10\% range for sources at
100\,Mpc. This is comparable to previous estimates of the calibration error in
the initial LIGO detectors. An interesting feature is that for distances
$\gtrsim 250$\,Mpc the upper extent of the uncertainty for H1 and L1 hits a
maximum at $\sim 25\%$, while the width of boxes narrows. This is due to our
``detectibility'' criteria, where for all distances we only see
those sources with a high enough \ac{SNR} that we would consider them
detectable, i.e.\ the weakest signals that could still be detected would
always have a single detector \ac{SNR} of $\sim 5.5$ no matter their distance, hence the
plateau. In addition there will also be fewer sources with \ac{SNR} higher than
this criteria at large distances, giving us a narrower range, and we
automatically exclude those with \ac{SNR} that are too small thus truncating
our uncertainty distribution at the upper end.  However, this does show that on
average for sources that are detectable out to 450\,Mpc we would be able to
constrain the calibration scale factor uncertainty for the \ac{aLIGO} detectors
to $\sim 20\%$. 

The largest \ac{SNR} contribution will generally come from the two
\ac{aLIGO} detectors and thus the detection criteria (\ac{SNR} threshold) will
not apply to the \ac{AdV} result. Hence, the \ac{SNR} in \ac{AdV} can be small
and thus the ability to constrain its scale factor becomes poor (although it
still provides information that the calibration is not grossly inaccurate). We
also see that true uncertainties achievable for recovering the calibration
scale factors are indeed very similar to the simple assumption that they would
be given by $\sim 1/{\rm SNR}$, i.e.\ at the upper end of the distribution the
\ac{SNR} in the two \ac{aLIGO} detectors will be $\sim 5.5$, which would be
expected to produce uncertainties of $\lesssim 20\%$. Our results are slightly
poorer than estimates based on this simple assumption due to correlations with
$\iota$ even within the constrained prior range, and slight correlations
between the scale factors for H1 and L1 as seen in Fig.~\ref{fig:bnspost}.

In Fig.~\ref{fig:nsbhresults} we see very similar results for the \ac{NSBH}
systems although the higher \ac{SNR} of the signals means that we can provide
$\sim 20\%\mbox{--}25\%$ uncertainties on the calibration scale factors for H1
and L1 out to greater distances. The distributions generally appear slightly
broader than those for the \ac{BNS} systems. This may be explained
by two factors. The first is from the fact that including the spin
parameter leads to strong correlations between the chirp mass, mass ratio and
spin. These strong correlations make convergence of the \ac{MCMC} take longer
and means we have fewer independent samples with which to estimate the posterior
distributions. This leads to a larger statistical fluctuation on the
results.  The second factor is that there appears to be a modest effect due to
the population of sources that is detectable. For distances at which the
\ac{BNS} and \ac{NSBH} systems would give the same percentage of observable
sources there is a slight increase in the mean and standard deviation of
\acp{SNR} of the \ac{NSBH} systems over the \ac{BNS} systems.

As noted previously, the mass distribution for black holes could be quite
different from the one used here, with masses extending to $\gtrsim
10$\,M$_{\odot}$. We expect these would produce similar results to those we see
in Fig.~\ref{fig:nsbhresults}, but again extending to further distances. However, at higher 
masses we would encounter the problem that the chirp mass will be less well constrained due to 
there being fewer observable cycles during the inspiral stage.
Including the merger and ring-down phase, which we currently ignore, will also
increase \ac{SNR} and act to reduce the calibration scale uncertainties, provided there is
sufficient \ac{SNR} in the inspiral to constrain the chirp mass.

\subsection{Parameter biases}

\begin{figure*}
\begin{center}

\begin{tabular}{c c}
\subcaptionbox{\label{subfig:pp50}}{
 \includegraphics[width=0.475\textwidth]{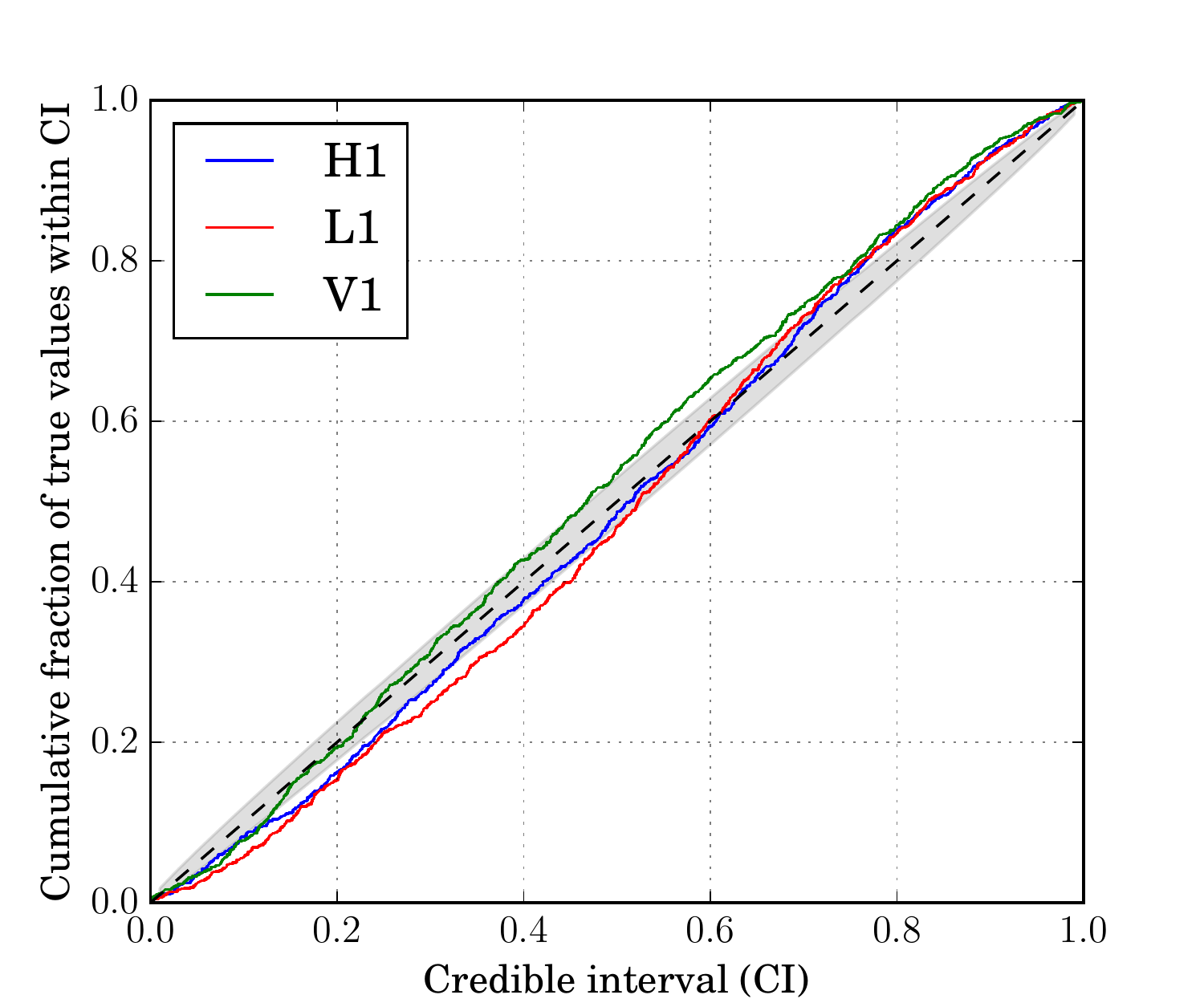}}

&

 \subcaptionbox{\label{subfig:pp500}}{
 \includegraphics[width=0.475\textwidth]{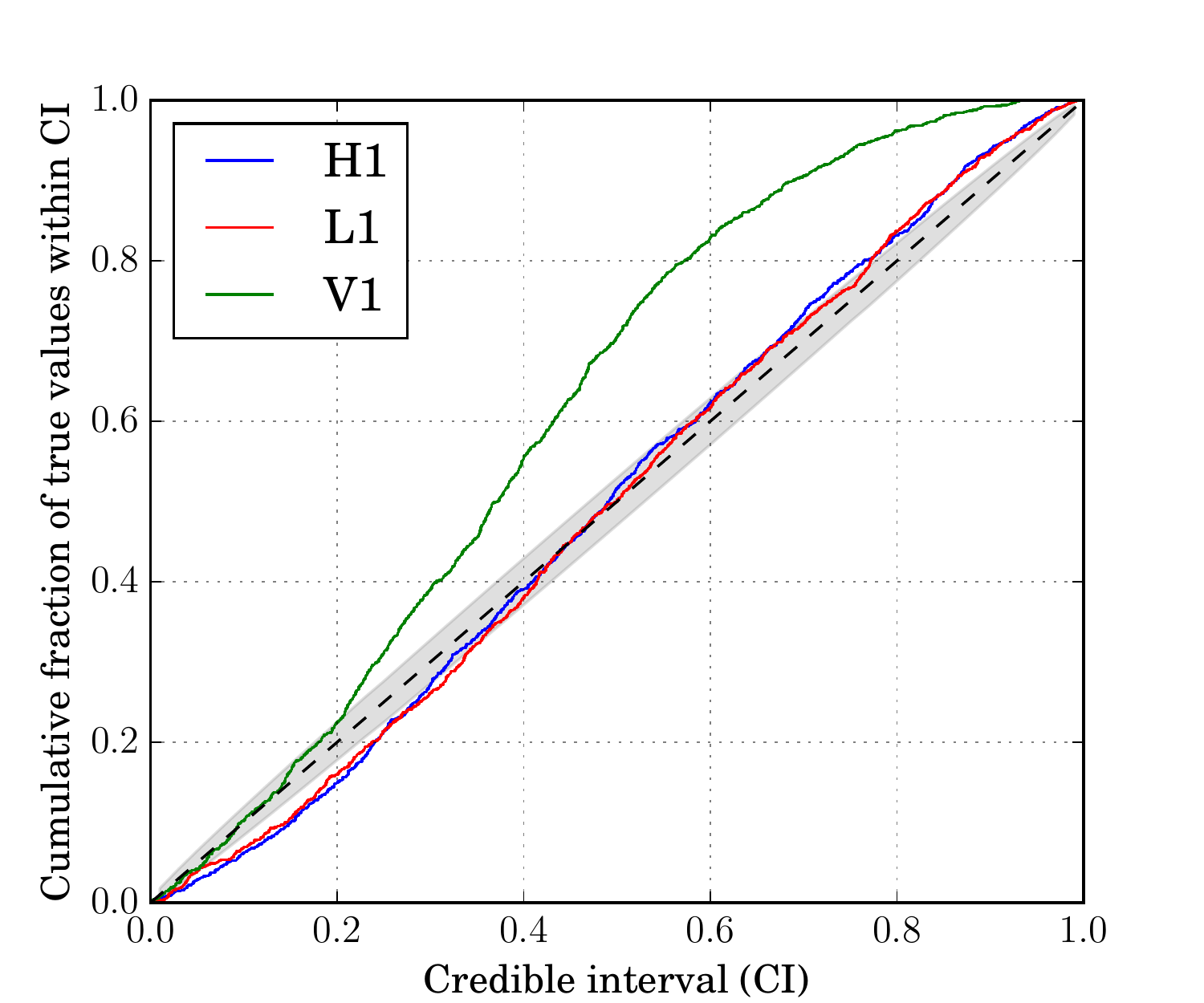}}
\end{tabular}
\end{center}
\caption{\label{fig:ppplots} Both figures show the cumulative fraction of true \scft values
found within a
given fractional credible interval versus the fractional credible interval for each detector for
the \ac{BNS} simulations at (\subref{subfig:pp50}) 50\,Mpc and (\subref{subfig:pp500})
500\,Mpc. The grey shaded region is a 95\% credible band for the expected deviations from diagonal.}
\end{figure*}

It is also interesting to see if our analysis has any bias on the recovered \scft \acp{PDF}, or whether they
are recovered in a way that is consistent with the simulated values.
To do this we have produced plots showing the cumulative fraction of the true (simulated)
\scft values recovered within given posterior credible intervals (see e.g.\ \cite{2014PhRvD..89h4060S} for the initial use
of this type of plot to check consistency of \ac{GW} analyes)
for the \ac{BNS} systems at 50 and 500\,Mpc (see Fig.~\ref{fig:ppplots}).
Self-consistent \acp{PDF} should show that a fraction of injected values falls within the corresponding credible interval,
and therefore the cumulative fraction should lie along the diagonal of the plot. Any
biases would show up as significant deviations from the diagonal.

Figure~\ref{subfig:pp50} shows
that for sources at 50\,Mpc there is no highly significant bias on the recovery of \scft for any of
the detectors. However, in Fig.~\ref{subfig:pp500}, with sources at 500\,Mpc, while
H1 and L1 follow the diagonal well, V1 goes significantly above the diagonal.
This shows that the \acp{PDF} on \scft for V1 are too broad. The reason for this is that at these
larger distances the \acp{PDF} on \scft for V1 are not well constrained and start to become dominated
by the prior. As we have previously noted this prior is not the same as the distribution used
to generate the \scft values, but is instead far broader and more conservative. So this bias is
showing up due to the simulated values of \scft being generated from a much more narrow range.

Despite the inconsistency in V1 it is satisfying to see that for the detectors for which the
\scft values are reasonably constrained (i.e.\ H1 and L1 in Fig.~\ref{fig:bnsresults}) the
actual value is recovered consistently, with little effect from the prior.

\section{Discussion\label{sec:discussion}}

We have established how well we can assess the detector calibration for
\ac{aLIGO} and \ac{AdV} using astrophysical sources as standard sirens. To do
this we require that a \ac{sGRB}, for which it has been possible to measure the
distance, is observed in coincidence with a \ac{CBC} signal in the \ac{GW}
detectors. This enables us to assume a known distance and sky position for the
source, and also limit the inclination of the source, which in turn allows us
to test the consistency of the detector calibration. We do this by including an
unknown scale factor on the true signal and noise, which we estimate given the
data from the three interferometer advanced detector network. We find that for
detectable \ac{BNS} sources the uncertainty on the calibration scale factor
could on average be determined to $\lesssim 10\%$ of its true value for the
\ac{aLIGO} and \ac{AdV} detectors out to 100\,Mpc. This is comparable to the
proposed accuracy level of the hardware calibration of the detectors. For
sources at the standard single detector \ac{BNS} horizon distance of $\sim
450$\,Mpc the scale factor could on average be determined to within $\lesssim
20\%$ of its true value for the \ac{aLIGO} detectors. Similar results were
found for \ac{NSBH} sources, although sources could be observed out to higher
distances.

The requirement of a coincident \ac{sGRB} with a known distance means that
there will likely be considerable delay (beyond initial \ac{GW} detection) for
the implementation of this method of calibration assessment.  Importantly
however, this method would provide an independent\footnote{There are caveats to this independence. 
We rely on a source having been detected, which most likely implies that calibration is not grossly 
inaccurate and is fairly consistent between detectors.} consistency check that the calibration is 
reasonably accurate in addition to the existing calibration methods \cite{2010NIMPA.624..223A}.
We acknowledge that we have made a number of simplifying assumptions, 
including that the waveforms we use accurately model the true signal, and that the overall 
calibration remains constant over the duration of a signal and its \ac{PSD} estimation. A further
major simplification made here is that the absolute calibration is constant
with frequency. Other authors (e.g.\ \cite{Vitale:2012,2013PhRvD..88h4044L}) have addressed
this issue in the context of robust parameter estimation for astrophysical
sources and in future work we intend to apply related methods for astrophysical
calibration.  We have also neglected any phase uncertainty in the calibration.

A further application of astrophysical calibration is the relative calibration
of \ac{GW} detectors.  In this case we are interested in obtaining
posterior distributions on the ratios of calibration scale factors and phases between
detectors. For \ac{CBC} events, in this scenario there would be no explicit
requirement for a redshift measurement and therefore a cosmologically inferred
distance. There would however be a requirement for an electromagnetically
inferred sky position of $O(\rm{degrees})$ accuracy. A more suitable source for
relative calibration estimation would be that of continuous \ac{GW} emission.
In that case, sky position and orientation information would be obtained to a
high level of accuracy from single detector analyses alone. A primary use for
relative calibration information would be to account for potential biases in
the sky position estimates for \ac{CBC} and burst sources (Vitale, Messenger \& Pitkin, in preparation). In this case,
amplitude information is a secondary, but important, factor in breaking
degeneracies on sky position estimates.    

We emphasize that our work has assumed that the \ac{PSD} used in the likelihood
function is an accurate representation of the \ac{PSD} at the time of the
observed \ac{CBC} signal. In reality the \ac{PSD} may be estimated from a
period close to, but not overlapping, the signal time and therefore may be
slightly different \cite{2013PhRvD..88h4044L}. Our result would therefore be
highly correlated with any uncertainty in the \ac{PSD} estimate, but would
still offer an upper limit on the calibration uncertainty. 

A further extension of this work would be for the calibration scale factor to
be estimated as a function of frequency. This may require the frequency series
to be divided into sections and the calibration scale factor estimated for
each. However, since the scale factor uncertainty is related to the \ac{SNR}
then the \ac{SNR} contribution in each frequency section is of relevance.
In this case Bayesian model selection may be applicable for selecting the
optimal number of frequency sections with which to estimate the scale factor.
The BayesLine algorithm \cite{2015PhRvD..91h4034L}, or the method described in
\cite{2013PhRvD..88h4044L}, may provide a natural way to perform such an
analysis. Indeed these methods, and those of \cite{Vitale:2012} are already
being used to marginalize over calibration uncertainties, but do not yet
attempt to estimate them.

It is also worth noting that astrophysical calibration may be possible for future planned 
space-based detectors such as eLISA \cite{2013GWN.....6....4A}. For such detectors there are 
several 
galactic binary systems known as ``verification'' binaries \cite{2006CQGra..23S.809S}, in that 
their masses and orbital parameters allow them to be guaranteed sources. These too could be used 
for assessing detector calibration. However, generally for these verification binaries 
the inclination of the system is not known and distances are fairly uncertain, so a precise 
calibration assessment may not be possible. However, there is at least one currently known 
eclipsing white dwarf binary (J0651) \cite{2041-8205-737-1-L23}, which does provide the system 
inclination \cite{2012ApJ...757L..21H, 2013ASPC..467...47K} and has an approximately 10\% distance 
uncertainty \cite{2041-8205-737-1-L23}, which makes it an excellent candidate for an astrophysical 
calibrator.

\acknowledgements

We acknowledge useful discussions within the LIGO-Virgo
Collaboration and specifically with Salvatore Vitale. L.W.\ was part funded for
this work through a Royal Astronomical Society Undergraduate Research Bursary.
M.P.\ and C.M.\ are funded by the Science and Technology Facilities Council (STFC) under Grant.
No.\ ST/L000946/1. C.M.\ is also supported by a Glasgow University Lord Kelvin
Adam Smith fellowship. We are grateful for computational resources provided by
Cardiff University, and funded by STFC Grant No.\ ST/I006285/1 supporting United Kingdom involvement in
the operation of Advanced LIGO. This document has LIGO DCC Report No. \dcc.

\bibliography{masterbib}

\begin{thebibliography}{43}%
\makeatletter
\providecommand \@ifxundefined [1]{%
 \@ifx{#1\undefined}
}%
\providecommand \@ifnum [1]{%
 \ifnum #1\expandafter \@firstoftwo
 \else \expandafter \@secondoftwo
 \fi
}%
\providecommand \@ifx [1]{%
 \ifx #1\expandafter \@firstoftwo
 \else \expandafter \@secondoftwo
 \fi
}%
\providecommand \natexlab [1]{#1}%
\providecommand \enquote  [1]{``#1''}%
\providecommand \bibnamefont  [1]{#1}%
\providecommand \bibfnamefont [1]{#1}%
\providecommand \citenamefont [1]{#1}%
\providecommand \href@noop [0]{\@secondoftwo}%
\providecommand \href [0]{\begingroup \@sanitize@url \@href}%
\providecommand \@href[1]{\@@startlink{#1}\@@href}%
\providecommand \@@href[1]{\endgroup#1\@@endlink}%
\providecommand \@sanitize@url [0]{\catcode `\\12\catcode `\$12\catcode
  `\&12\catcode `\#12\catcode `\^12\catcode `\_12\catcode `\%12\relax}%
\providecommand \@@startlink[1]{}%
\providecommand \@@endlink[0]{}%
\providecommand \url  [0]{\begingroup\@sanitize@url \@url }%
\providecommand \@url [1]{\endgroup\@href {#1}{\urlprefix }}%
\providecommand \urlprefix  [0]{URL }%
\providecommand \Eprint [0]{\href }%
\providecommand \doibase [0]{http://dx.doi.org/}%
\providecommand \selectlanguage [0]{\@gobble}%
\providecommand \bibinfo  [0]{\@secondoftwo}%
\providecommand \bibfield  [0]{\@secondoftwo}%
\providecommand \translation [1]{[#1]}%
\providecommand \BibitemOpen [0]{}%
\providecommand \bibitemStop [0]{}%
\providecommand \bibitemNoStop [0]{.\EOS\space}%
\providecommand \EOS [0]{\spacefactor3000\relax}%
\providecommand \BibitemShut  [1]{\csname bibitem#1\endcsname}%
\let\auto@bib@innerbib\@empty
\bibitem [{\citenamefont {{Abadie}}\ \emph
  {et~al.}(2010{\natexlab{a}})\citenamefont {{Abadie}} \emph
  {et~al.}}]{2010CQGra..27q3001A}%
  \BibitemOpen
  \bibfield  {author} {\bibinfo {author} {\bibfnamefont {J.}~\bibnamefont
  {{Abadie}}} \emph {et~al.},\ }\href {\doibase 10.1088/0264-9381/27/17/173001}
  {\bibfield  {journal} {\bibinfo  {journal} {Class. Quantum Grav.}\ }\textbf
  {\bibinfo {volume} {27}},\ \bibinfo {eid} {173001} (\bibinfo {year}
  {2010}{\natexlab{a}})},\ \Eprint {http://arxiv.org/abs/1003.2480}
  {arXiv:1003.2480 [astro-ph.HE]} \BibitemShut {NoStop}%
\bibitem [{\citenamefont {Schutz}(1986)}]{1986Natur.323..310S}%
  \BibitemOpen
  \bibfield  {author} {\bibinfo {author} {\bibfnamefont {B.~F.}\ \bibnamefont
  {Schutz}},\ }\href@noop {} {\bibfield  {journal} {\bibinfo  {journal}
  {Nature}\ }\textbf {\bibinfo {volume} {323}},\ \bibinfo {pages} {310}
  (\bibinfo {year} {1986})}\BibitemShut {NoStop}%
\bibitem [{\citenamefont {{Narayan}}\ \emph {et~al.}(1992)\citenamefont
  {{Narayan}}, \citenamefont {{Paczynski}},\ and\ \citenamefont
  {{Piran}}}]{1992ApJ...395L..83N}%
  \BibitemOpen
  \bibfield  {author} {\bibinfo {author} {\bibfnamefont {R.}~\bibnamefont
  {{Narayan}}}, \bibinfo {author} {\bibfnamefont {B.}~\bibnamefont
  {{Paczynski}}}, \ and\ \bibinfo {author} {\bibfnamefont {T.}~\bibnamefont
  {{Piran}}},\ }\href {\doibase 10.1086/186493} {\bibfield  {journal} {\bibinfo
   {journal} {\apj}\ }\textbf {\bibinfo {volume} {395}},\ \bibinfo {pages}
  {L83} (\bibinfo {year} {1992})},\ \Eprint
  {http://arxiv.org/abs/astro-ph/9204001} {astro-ph/9204001} \BibitemShut
  {NoStop}%
\bibitem [{\citenamefont {Coward}\ \emph {et~al.}(2012)\citenamefont {Coward},
  \citenamefont {Howell}, \citenamefont {Piran}, \citenamefont {Stratta},
  \citenamefont {Branchesi}, \citenamefont {Bromberg}, \citenamefont {Gendre},
  \citenamefont {Burman},\ and\ \citenamefont {Guetta}}]{2012MNRAS.425.2668C}%
  \BibitemOpen
  \bibfield  {author} {\bibinfo {author} {\bibfnamefont {D.~M.}\ \bibnamefont
  {Coward}}, \bibinfo {author} {\bibfnamefont {E.~J.}\ \bibnamefont {Howell}},
  \bibinfo {author} {\bibfnamefont {T.}~\bibnamefont {Piran}}, \bibinfo
  {author} {\bibfnamefont {G.}~\bibnamefont {Stratta}}, \bibinfo {author}
  {\bibfnamefont {M.}~\bibnamefont {Branchesi}}, \bibinfo {author}
  {\bibfnamefont {O.}~\bibnamefont {Bromberg}}, \bibinfo {author}
  {\bibfnamefont {B.}~\bibnamefont {Gendre}}, \bibinfo {author} {\bibfnamefont
  {R.~R.}\ \bibnamefont {Burman}}, \ and\ \bibinfo {author} {\bibfnamefont
  {D.}~\bibnamefont {Guetta}},\ }\href@noop {} {\bibfield  {journal} {\bibinfo
  {journal} {Mon. Not. R. Astron. Soc.}\ }\textbf {\bibinfo {volume} {425}},\
  \bibinfo {pages} {2668} (\bibinfo {year} {2012})}\BibitemShut {NoStop}%
\bibitem [{\citenamefont {{Kim}}\ \emph {et~al.}(2015)\citenamefont {{Kim}},
  \citenamefont {{Perera}},\ and\ \citenamefont
  {{McLaughlin}}}]{2015MNRAS.448..928K}%
  \BibitemOpen
  \bibfield  {author} {\bibinfo {author} {\bibfnamefont {C.}~\bibnamefont
  {{Kim}}}, \bibinfo {author} {\bibfnamefont {B.~B.~P.}\ \bibnamefont
  {{Perera}}}, \ and\ \bibinfo {author} {\bibfnamefont {M.~A.}\ \bibnamefont
  {{McLaughlin}}},\ }\href {\doibase 10.1093/mnras/stu2729} {\bibfield
  {journal} {\bibinfo  {journal} {Mon. Not. R. Astron. Soc.}\ }\textbf
  {\bibinfo {volume} {448}},\ \bibinfo {pages} {928} (\bibinfo {year}
  {2015})},\ \Eprint {http://arxiv.org/abs/1308.4676} {arXiv:1308.4676
  [astro-ph.SR]} \BibitemShut {NoStop}%
\bibitem [{\citenamefont {Berger}\ \emph {et~al.}(2007)\citenamefont {Berger}
  \emph {et~al.}}]{0004-637X-664-2-1000}%
  \BibitemOpen
  \bibfield  {author} {\bibinfo {author} {\bibfnamefont {E.}~\bibnamefont
  {Berger}} \emph {et~al.},\ }\href
  {http://stacks.iop.org/0004-637X/664/i=2/a=1000} {\bibfield  {journal}
  {\bibinfo  {journal} {\apj}\ }\textbf {\bibinfo {volume} {664}},\ \bibinfo
  {pages} {1000} (\bibinfo {year} {2007})}\BibitemShut {NoStop}%
\bibitem [{\citenamefont {{Planck Collaboration}}\ \emph
  {et~al.}(2015)\citenamefont {{Planck Collaboration}}, \citenamefont {{Ade}}
  \emph {et~al.}}]{2015arXiv150201589P}%
  \BibitemOpen
  \bibfield  {author} {\bibinfo {author} {\bibnamefont {{Planck
  Collaboration}}}, \bibinfo {author} {\bibfnamefont {P.~A.~R.}\ \bibnamefont
  {{Ade}}},  \emph {et~al.},\ }\href@noop {} {\bibfield  {journal} {\bibinfo
  {journal} {ArXiv e-prints}\ } (\bibinfo {year} {2015})},\ \Eprint
  {http://arxiv.org/abs/1502.01589} {arXiv:1502.01589} \BibitemShut {NoStop}%
\bibitem [{\citenamefont {{Abadie}}\ \emph
  {et~al.}(2010{\natexlab{b}})\citenamefont {{Abadie}} \emph
  {et~al.}}]{2010NIMPA.624..223A}%
  \BibitemOpen
  \bibfield  {author} {\bibinfo {author} {\bibfnamefont {J.}~\bibnamefont
  {{Abadie}}} \emph {et~al.},\ }\href {\doibase 10.1016/j.nima.2010.07.089}
  {\bibfield  {journal} {\bibinfo  {journal} {NIMPA}\ }\textbf {\bibinfo
  {volume} {624}},\ \bibinfo {pages} {223} (\bibinfo {year}
  {2010}{\natexlab{b}})},\ \Eprint {http://arxiv.org/abs/1007.3973}
  {arXiv:1007.3973 [gr-qc]} \BibitemShut {NoStop}%
\bibitem [{\citenamefont {{Vitale}}\ \emph {et~al.}(2012)\citenamefont
  {{Vitale}}, \citenamefont {{Del Pozzo}}, \citenamefont {{Li}}, \citenamefont
  {{Van Den Broeck}}, \citenamefont {{Mandel}}, \citenamefont {{Aylott}},\ and\
  \citenamefont {{Veitch}}}]{Vitale:2012}%
  \BibitemOpen
  \bibfield  {author} {\bibinfo {author} {\bibfnamefont {S.}~\bibnamefont
  {{Vitale}}}, \bibinfo {author} {\bibfnamefont {W.}~\bibnamefont {{Del
  Pozzo}}}, \bibinfo {author} {\bibfnamefont {T.~G.~F.}\ \bibnamefont {{Li}}},
  \bibinfo {author} {\bibfnamefont {C.~V.}\ \bibnamefont {{Van Den Broeck}}},
  \bibinfo {author} {\bibfnamefont {I.}~\bibnamefont {{Mandel}}}, \bibinfo
  {author} {\bibfnamefont {B.}~\bibnamefont {{Aylott}}}, \ and\ \bibinfo
  {author} {\bibfnamefont {J.}~\bibnamefont {{Veitch}}},\ }\href {\doibase
  10.1103/PhysRevD.85.064034} {\bibfield  {journal} {\bibinfo  {journal}
  {\prd}\ }\textbf {\bibinfo {volume} {85}},\ \bibinfo {eid} {064034} (\bibinfo
  {year} {2012})},\ \Eprint {http://arxiv.org/abs/1111.3044} {arXiv:1111.3044
  [gr-qc]} \BibitemShut {NoStop}%
\bibitem [{\citenamefont {{Littenberg}}\ \emph {et~al.}(2013)\citenamefont
  {{Littenberg}}, \citenamefont {{Coughlin}}, \citenamefont {{Farr}},\ and\
  \citenamefont {{Farr}}}]{2013PhRvD..88h4044L}%
  \BibitemOpen
  \bibfield  {author} {\bibinfo {author} {\bibfnamefont {T.~B.}\ \bibnamefont
  {{Littenberg}}}, \bibinfo {author} {\bibfnamefont {M.}~\bibnamefont
  {{Coughlin}}}, \bibinfo {author} {\bibfnamefont {B.}~\bibnamefont {{Farr}}},
  \ and\ \bibinfo {author} {\bibfnamefont {W.~M.}\ \bibnamefont {{Farr}}},\
  }\href {\doibase 10.1103/PhysRevD.88.084044} {\bibfield  {journal} {\bibinfo
  {journal} {\prd}\ }\textbf {\bibinfo {volume} {88}},\ \bibinfo {eid} {084044}
  (\bibinfo {year} {2013})},\ \Eprint {http://arxiv.org/abs/1307.8195}
  {arXiv:1307.8195 [astro-ph.IM]} \BibitemShut {NoStop}%
\bibitem [{\citenamefont {{Singer}}\ \emph {et~al.}(2014)\citenamefont
  {{Singer}} \emph {et~al.}}]{2014ApJ...795..105S}%
  \BibitemOpen
  \bibfield  {author} {\bibinfo {author} {\bibfnamefont {L.~P.}\ \bibnamefont
  {{Singer}}} \emph {et~al.},\ }\href {\doibase 10.1088/0004-637X/795/2/105}
  {\bibfield  {journal} {\bibinfo  {journal} {\apj}\ }\textbf {\bibinfo
  {volume} {795}},\ \bibinfo {eid} {105} (\bibinfo {year} {2014})},\ \Eprint
  {http://arxiv.org/abs/1404.5623} {arXiv:1404.5623 [astro-ph.HE]} \BibitemShut
  {NoStop}%
\bibitem [{\citenamefont {Sathyaprakash}\ \emph {et~al.}(2010)\citenamefont
  {Sathyaprakash}, \citenamefont {Schutz},\ and\ \citenamefont {van~den
  Broeck}}]{2010CQGra..27u5006S}%
  \BibitemOpen
  \bibfield  {author} {\bibinfo {author} {\bibfnamefont {B.~S.}\ \bibnamefont
  {Sathyaprakash}}, \bibinfo {author} {\bibfnamefont {B.~F.}\ \bibnamefont
  {Schutz}}, \ and\ \bibinfo {author} {\bibfnamefont {C.}~\bibnamefont {van~den
  Broeck}},\ }\href@noop {} {\bibfield  {journal} {\bibinfo  {journal} {Class.
  Quantum Grav.}\ }\textbf {\bibinfo {volume} {27}},\ \bibinfo {pages} {215006}
  (\bibinfo {year} {2010})}\BibitemShut {NoStop}%
\bibitem [{\citenamefont {Zhao}\ \emph {et~al.}(2011)\citenamefont {Zhao},
  \citenamefont {Van Den~Broeck}, \citenamefont {Baskaran},\ and\ \citenamefont
  {Li}}]{2011PhRvD..83b3005Z}%
  \BibitemOpen
  \bibfield  {author} {\bibinfo {author} {\bibfnamefont {W.}~\bibnamefont
  {Zhao}}, \bibinfo {author} {\bibfnamefont {C.}~\bibnamefont {Van
  Den~Broeck}}, \bibinfo {author} {\bibfnamefont {D.}~\bibnamefont {Baskaran}},
  \ and\ \bibinfo {author} {\bibfnamefont {T.~G.~F.}\ \bibnamefont {Li}},\
  }\href@noop {} {\bibfield  {journal} {\bibinfo  {journal} {\prd}\ }\textbf
  {\bibinfo {volume} {83}},\ \bibinfo {pages} {023005} (\bibinfo {year}
  {2011})}\BibitemShut {NoStop}%
\bibitem [{\citenamefont {{Punturo}}\ \emph {et~al.}(2010)\citenamefont
  {{Punturo}} \emph {et~al.}}]{2010CQGra..27h4007P}%
  \BibitemOpen
  \bibfield  {author} {\bibinfo {author} {\bibfnamefont {M.}~\bibnamefont
  {{Punturo}}} \emph {et~al.},\ }\href {\doibase 10.1088/0264-9381/27/8/084007}
  {\bibfield  {journal} {\bibinfo  {journal} {Classical and Quantum Gravity}\
  }\textbf {\bibinfo {volume} {27}},\ \bibinfo {eid} {084007} (\bibinfo {year}
  {2010})}\BibitemShut {NoStop}%
\bibitem [{\citenamefont {Taylor}\ \emph {et~al.}(2012)\citenamefont {Taylor},
  \citenamefont {Gair},\ and\ \citenamefont {Mandel}}]{2012PhRvD..85b3535T}%
  \BibitemOpen
  \bibfield  {author} {\bibinfo {author} {\bibfnamefont {S.~R.}\ \bibnamefont
  {Taylor}}, \bibinfo {author} {\bibfnamefont {J.~R.}\ \bibnamefont {Gair}}, \
  and\ \bibinfo {author} {\bibfnamefont {I.}~\bibnamefont {Mandel}},\
  }\href@noop {} {\bibfield  {journal} {\bibinfo  {journal} {\prd}\ }\textbf
  {\bibinfo {volume} {85}},\ \bibinfo {pages} {023535} (\bibinfo {year}
  {2012})}\BibitemShut {NoStop}%
\bibitem [{\citenamefont {{Finn}}(1996)}]{1996PhRvD..53.2878F}%
  \BibitemOpen
  \bibfield  {author} {\bibinfo {author} {\bibfnamefont {L.~S.}\ \bibnamefont
  {{Finn}}},\ }\href {\doibase 10.1103/PhysRevD.53.2878} {\bibfield  {journal}
  {\bibinfo  {journal} {\prd}\ }\textbf {\bibinfo {volume} {53}},\ \bibinfo
  {pages} {2878} (\bibinfo {year} {1996})},\ \Eprint
  {http://arxiv.org/abs/gr-qc/9601048} {gr-qc/9601048} \BibitemShut {NoStop}%
\bibitem [{\citenamefont {{Taylor}}\ and\ \citenamefont
  {{Gair}}(2012)}]{2012PhRvD..86b3502T}%
  \BibitemOpen
  \bibfield  {author} {\bibinfo {author} {\bibfnamefont {S.~R.}\ \bibnamefont
  {{Taylor}}}\ and\ \bibinfo {author} {\bibfnamefont {J.~R.}\ \bibnamefont
  {{Gair}}},\ }\href {\doibase 10.1103/PhysRevD.86.023502} {\bibfield
  {journal} {\bibinfo  {journal} {\prd}\ }\textbf {\bibinfo {volume} {86}},\
  \bibinfo {eid} {023502} (\bibinfo {year} {2012})},\ \Eprint
  {http://arxiv.org/abs/1204.6739} {arXiv:1204.6739 [astro-ph.CO]} \BibitemShut
  {NoStop}%
\bibitem [{\citenamefont {{Messenger}}\ and\ \citenamefont
  {{Read}}(2012)}]{Messenger:2011ux}%
  \BibitemOpen
  \bibfield  {author} {\bibinfo {author} {\bibfnamefont {C.}~\bibnamefont
  {{Messenger}}}\ and\ \bibinfo {author} {\bibfnamefont {J.}~\bibnamefont
  {{Read}}},\ }\href {\doibase 10.1103/PhysRevLett.108.091101} {\bibfield
  {journal} {\bibinfo  {journal} {\prl}\ }\textbf {\bibinfo {volume} {108}},\
  \bibinfo {eid} {091101} (\bibinfo {year} {2012})},\ \Eprint
  {http://arxiv.org/abs/1107.5725} {arXiv:1107.5725 [gr-qc]} \BibitemShut
  {NoStop}%
\bibitem [{\citenamefont {{Messenger}}\ \emph {et~al.}(2014)\citenamefont
  {{Messenger}}, \citenamefont {{Takami}}, \citenamefont {{Gossan}},
  \citenamefont {{Rezzolla}},\ and\ \citenamefont
  {{Sathyaprakash}}}]{2013arXiv1312.1862M}%
  \BibitemOpen
  \bibfield  {author} {\bibinfo {author} {\bibfnamefont {C.}~\bibnamefont
  {{Messenger}}}, \bibinfo {author} {\bibfnamefont {K.}~\bibnamefont
  {{Takami}}}, \bibinfo {author} {\bibfnamefont {S.}~\bibnamefont {{Gossan}}},
  \bibinfo {author} {\bibfnamefont {L.}~\bibnamefont {{Rezzolla}}}, \ and\
  \bibinfo {author} {\bibfnamefont {B.~S.}\ \bibnamefont {{Sathyaprakash}}},\
  }\href {\doibase 10.1103/PhysRevX.4.041004} {\bibfield  {journal} {\bibinfo
  {journal} {Phys. Rev. X}\ }\textbf {\bibinfo {volume} {4}},\ \bibinfo {eid}
  {041004} (\bibinfo {year} {2014})},\ \Eprint {http://arxiv.org/abs/1312.1862}
  {arXiv:1312.1862 [gr-qc]} \BibitemShut {NoStop}%
\bibitem [{\citenamefont {Eichler}\ \emph {et~al.}(1989)\citenamefont
  {Eichler}, \citenamefont {Livio}, \citenamefont {Piran},\ and\ \citenamefont
  {Schramm}}]{1989Natur.340..126E}%
  \BibitemOpen
  \bibfield  {author} {\bibinfo {author} {\bibfnamefont {D.}~\bibnamefont
  {Eichler}}, \bibinfo {author} {\bibfnamefont {M.}~\bibnamefont {Livio}},
  \bibinfo {author} {\bibfnamefont {T.}~\bibnamefont {Piran}}, \ and\ \bibinfo
  {author} {\bibfnamefont {D.~N.}\ \bibnamefont {Schramm}},\ }\href@noop {}
  {\bibfield  {journal} {\bibinfo  {journal} {Nature}\ }\textbf {\bibinfo
  {volume} {340}},\ \bibinfo {pages} {126} (\bibinfo {year}
  {1989})}\BibitemShut {NoStop}%
\bibitem [{\citenamefont {{Gehrels}}\ \emph {et~al.}(2005)\citenamefont
  {{Gehrels}} \emph {et~al.}}]{2005Natur.437..851G}%
  \BibitemOpen
  \bibfield  {author} {\bibinfo {author} {\bibfnamefont {N.}~\bibnamefont
  {{Gehrels}}} \emph {et~al.},\ }\href {\doibase 10.1038/nature04142}
  {\bibfield  {journal} {\bibinfo  {journal} {Nature}\ }\textbf {\bibinfo
  {volume} {437}},\ \bibinfo {pages} {851} (\bibinfo {year} {2005})},\ \Eprint
  {http://arxiv.org/abs/astro-ph/0505630} {astro-ph/0505630} \BibitemShut
  {NoStop}%
\bibitem [{\citenamefont {{Castro-Tirado}}\ \emph {et~al.}(2015)\citenamefont
  {{Castro-Tirado}}, \citenamefont {{Sanchez-Ramirez}}, \citenamefont
  {{Gorosabel}},\ and\ \citenamefont {{Scarpa}}}]{2015GCN..17278...1C}%
  \BibitemOpen
  \bibfield  {author} {\bibinfo {author} {\bibfnamefont {A.~J.}\ \bibnamefont
  {{Castro-Tirado}}}, \bibinfo {author} {\bibfnamefont {R.}~\bibnamefont
  {{Sanchez-Ramirez}}}, \bibinfo {author} {\bibfnamefont {J.}~\bibnamefont
  {{Gorosabel}}}, \ and\ \bibinfo {author} {\bibfnamefont {R.}~\bibnamefont
  {{Scarpa}}},\ }\href@noop {} {\bibfield  {journal} {\bibinfo  {journal} {GRB
  Coordinates Network}\ }\textbf {\bibinfo {volume} {17278}},\ \bibinfo {pages}
  {1} (\bibinfo {year} {2015})}\BibitemShut {NoStop}%
\bibitem [{\citenamefont {{Abbott}}\ \emph {et~al.}(2016)\citenamefont
  {{Abbott}}, \citenamefont {{Abbott}}, \citenamefont {{Abbott}}, \citenamefont
  {{Abernathy}}, \citenamefont {{Acernese}}, \citenamefont {{Ackley}},
  \citenamefont {{Adams}}, \citenamefont {{Adams}}, \citenamefont {{Addesso}},
  \citenamefont {{Adhikari}},\ and\ \citenamefont
  {et~al.}}]{2013arXiv1304.0670L}%
  \BibitemOpen
  \bibfield  {author} {\bibinfo {author} {\bibfnamefont {B.~P.}\ \bibnamefont
  {{Abbott}}}, \bibinfo {author} {\bibfnamefont {R.}~\bibnamefont {{Abbott}}},
  \bibinfo {author} {\bibfnamefont {T.~D.}\ \bibnamefont {{Abbott}}}, \bibinfo
  {author} {\bibfnamefont {M.~R.}\ \bibnamefont {{Abernathy}}}, \bibinfo
  {author} {\bibfnamefont {F.}~\bibnamefont {{Acernese}}}, \bibinfo {author}
  {\bibfnamefont {K.}~\bibnamefont {{Ackley}}}, \bibinfo {author}
  {\bibfnamefont {C.}~\bibnamefont {{Adams}}}, \bibinfo {author} {\bibfnamefont
  {T.}~\bibnamefont {{Adams}}}, \bibinfo {author} {\bibfnamefont
  {P.}~\bibnamefont {{Addesso}}}, \bibinfo {author} {\bibfnamefont {R.~X.}\
  \bibnamefont {{Adhikari}}}, \ and\ \bibinfo {author} {\bibnamefont
  {et~al.}},\ }\href {\doibase 10.1007/lrr-2016-1} {\bibfield  {journal}
  {\bibinfo  {journal} {Living Reviews in Relativity}\ }\textbf {\bibinfo
  {volume} {19}} (\bibinfo {year} {2016}),\ 10.1007/lrr-2016-1},\ \Eprint
  {http://arxiv.org/abs/1304.0670} {arXiv:1304.0670 [gr-qc]} \BibitemShut
  {NoStop}%
\bibitem [{\citenamefont {Chen}\ and\ \citenamefont
  {Holz}(2013)}]{2013PhRvL.111r1101C}%
  \BibitemOpen
  \bibfield  {author} {\bibinfo {author} {\bibfnamefont {H.-Y.}\ \bibnamefont
  {Chen}}\ and\ \bibinfo {author} {\bibfnamefont {D.~E.}\ \bibnamefont
  {Holz}},\ }\href@noop {} {\bibfield  {journal} {\bibinfo  {journal} {\prl}\
  }\textbf {\bibinfo {volume} {111}},\ \bibinfo {pages} {181101} (\bibinfo
  {year} {2013})}\BibitemShut {NoStop}%
\bibitem [{\citenamefont {{Kelley}}\ \emph {et~al.}(2013)\citenamefont
  {{Kelley}}, \citenamefont {{Mandel}},\ and\ \citenamefont
  {{Ramirez-Ruiz}}}]{Kelley:2012fl}%
  \BibitemOpen
  \bibfield  {author} {\bibinfo {author} {\bibfnamefont {L.~Z.}\ \bibnamefont
  {{Kelley}}}, \bibinfo {author} {\bibfnamefont {I.}~\bibnamefont {{Mandel}}},
  \ and\ \bibinfo {author} {\bibfnamefont {E.}~\bibnamefont {{Ramirez-Ruiz}}},\
  }\href {\doibase 10.1103/PhysRevD.87.123004} {\bibfield  {journal} {\bibinfo
  {journal} {\prd}\ }\textbf {\bibinfo {volume} {87}},\ \bibinfo {eid} {123004}
  (\bibinfo {year} {2013})},\ \Eprint {http://arxiv.org/abs/1209.3027}
  {arXiv:1209.3027 [astro-ph.IM]} \BibitemShut {NoStop}%
\bibitem [{\citenamefont {{Siellez}}\ \emph {et~al.}(2014)\citenamefont
  {{Siellez}}, \citenamefont {{Bo{\"e}r}},\ and\ \citenamefont
  {{Gendre}}}]{2014arXiv1405.2254S}%
  \BibitemOpen
  \bibfield  {author} {\bibinfo {author} {\bibfnamefont {K.}~\bibnamefont
  {{Siellez}}}, \bibinfo {author} {\bibfnamefont {M.}~\bibnamefont
  {{Bo{\"e}r}}}, \ and\ \bibinfo {author} {\bibfnamefont {B.}~\bibnamefont
  {{Gendre}}},\ }\href {\doibase 10.1093/mnras/stt1915} {\bibfield  {journal}
  {\bibinfo  {journal} {Mon. Not. R. Astron. Soc.}\ }\textbf {\bibinfo {volume}
  {437}},\ \bibinfo {pages} {649} (\bibinfo {year} {2014})},\ \Eprint
  {http://arxiv.org/abs/1310.2106} {arXiv:1310.2106 [astro-ph.HE]} \BibitemShut
  {NoStop}%
\bibitem [{\citenamefont {{Wanderman}}\ and\ \citenamefont
  {{Piran}}(2015)}]{2015MNRAS.448.3026W}%
  \BibitemOpen
  \bibfield  {author} {\bibinfo {author} {\bibfnamefont {D.}~\bibnamefont
  {{Wanderman}}}\ and\ \bibinfo {author} {\bibfnamefont {T.}~\bibnamefont
  {{Piran}}},\ }\href {\doibase 10.1093/mnras/stv123} {\bibfield  {journal}
  {\bibinfo  {journal} {Mon. Not. R. Astron. Soc.}\ }\textbf {\bibinfo {volume}
  {448}},\ \bibinfo {pages} {3026} (\bibinfo {year} {2015})},\ \Eprint
  {http://arxiv.org/abs/1405.5878} {arXiv:1405.5878 [astro-ph.HE]} \BibitemShut
  {NoStop}%
\bibitem [{\citenamefont {{Blackburn}}\ \emph {et~al.}(2015)\citenamefont
  {{Blackburn}}, \citenamefont {{Briggs}}, \citenamefont {{Camp}},
  \citenamefont {{Christensen}}, \citenamefont {{Connaughton}}, \citenamefont
  {{Jenke}}, \citenamefont {{Remillard}},\ and\ \citenamefont
  {{Veitch}}}]{2015ApJS..217....8B}%
  \BibitemOpen
  \bibfield  {author} {\bibinfo {author} {\bibfnamefont {L.}~\bibnamefont
  {{Blackburn}}}, \bibinfo {author} {\bibfnamefont {M.~S.}\ \bibnamefont
  {{Briggs}}}, \bibinfo {author} {\bibfnamefont {J.}~\bibnamefont {{Camp}}},
  \bibinfo {author} {\bibfnamefont {N.}~\bibnamefont {{Christensen}}}, \bibinfo
  {author} {\bibfnamefont {V.}~\bibnamefont {{Connaughton}}}, \bibinfo {author}
  {\bibfnamefont {P.}~\bibnamefont {{Jenke}}}, \bibinfo {author} {\bibfnamefont
  {R.~A.}\ \bibnamefont {{Remillard}}}, \ and\ \bibinfo {author} {\bibfnamefont
  {J.}~\bibnamefont {{Veitch}}},\ }\href {\doibase 10.1088/0067-0049/217/1/8}
  {\bibfield  {journal} {\bibinfo  {journal} {Astrophys. J. Suppl. S.}\
  }\textbf {\bibinfo {volume} {217}},\ \bibinfo {eid} {8} (\bibinfo {year}
  {2015})},\ \Eprint {http://arxiv.org/abs/1410.0929} {arXiv:1410.0929
  [astro-ph.HE]} \BibitemShut {NoStop}%
\bibitem [{\citenamefont {{Fairhurst}}(2011)}]{2011CQGra..28j5021F}%
  \BibitemOpen
  \bibfield  {author} {\bibinfo {author} {\bibfnamefont {S.}~\bibnamefont
  {{Fairhurst}}},\ }\href {\doibase 10.1088/0264-9381/28/10/105021} {\bibfield
  {journal} {\bibinfo  {journal} {Class. Quantum Grav.}\ }\textbf {\bibinfo
  {volume} {28}},\ \bibinfo {eid} {105021} (\bibinfo {year} {2011})},\ \Eprint
  {http://arxiv.org/abs/1010.6192} {arXiv:1010.6192 [gr-qc]} \BibitemShut
  {NoStop}%
\bibitem [{\citenamefont {{Abadie}}\ \emph
  {et~al.}(2012{\natexlab{a}})\citenamefont {{Abadie}} \emph
  {et~al.}}]{2012ApJ...760...12A}%
  \BibitemOpen
  \bibfield  {author} {\bibinfo {author} {\bibfnamefont {J.}~\bibnamefont
  {{Abadie}}} \emph {et~al.},\ }\href {\doibase 10.1088/0004-637X/760/1/12}
  {\bibfield  {journal} {\bibinfo  {journal} {\apj}\ }\textbf {\bibinfo
  {volume} {760}},\ \bibinfo {eid} {12} (\bibinfo {year}
  {2012}{\natexlab{a}})},\ \Eprint {http://arxiv.org/abs/1205.2216}
  {arXiv:1205.2216 [astro-ph.HE]} \BibitemShut {NoStop}%
\bibitem [{\citenamefont {{Buonanno}}\ \emph {et~al.}(2009)\citenamefont
  {{Buonanno}}, \citenamefont {{Iyer}}, \citenamefont {{Ochsner}},
  \citenamefont {{Pan}},\ and\ \citenamefont
  {{Sathyaprakash}}}]{2009PhRvD..80h4043B}%
  \BibitemOpen
  \bibfield  {author} {\bibinfo {author} {\bibfnamefont {A.}~\bibnamefont
  {{Buonanno}}}, \bibinfo {author} {\bibfnamefont {B.~R.}\ \bibnamefont
  {{Iyer}}}, \bibinfo {author} {\bibfnamefont {E.}~\bibnamefont {{Ochsner}}},
  \bibinfo {author} {\bibfnamefont {Y.}~\bibnamefont {{Pan}}}, \ and\ \bibinfo
  {author} {\bibfnamefont {B.~S.}\ \bibnamefont {{Sathyaprakash}}},\ }\href
  {\doibase 10.1103/PhysRevD.80.084043} {\bibfield  {journal} {\bibinfo
  {journal} {\prd}\ }\textbf {\bibinfo {volume} {80}},\ \bibinfo {eid} {084043}
  (\bibinfo {year} {2009})},\ \Eprint {http://arxiv.org/abs/0907.0700}
  {arXiv:0907.0700 [gr-qc]} \BibitemShut {NoStop}%
\bibitem [{\citenamefont {{Brown}}\ \emph {et~al.}(2012)\citenamefont
  {{Brown}}, \citenamefont {{Harry}}, \citenamefont {{Lundgren}},\ and\
  \citenamefont {{Nitz}}}]{2012PhRvD..86h4017B}%
  \BibitemOpen
  \bibfield  {author} {\bibinfo {author} {\bibfnamefont {D.~A.}\ \bibnamefont
  {{Brown}}}, \bibinfo {author} {\bibfnamefont {I.}~\bibnamefont {{Harry}}},
  \bibinfo {author} {\bibfnamefont {A.}~\bibnamefont {{Lundgren}}}, \ and\
  \bibinfo {author} {\bibfnamefont {A.~H.}\ \bibnamefont {{Nitz}}},\ }\href
  {\doibase 10.1103/PhysRevD.86.084017} {\bibfield  {journal} {\bibinfo
  {journal} {\prd}\ }\textbf {\bibinfo {volume} {86}},\ \bibinfo {eid} {084017}
  (\bibinfo {year} {2012})},\ \Eprint {http://arxiv.org/abs/1207.6406}
  {arXiv:1207.6406 [gr-qc]} \BibitemShut {NoStop}%
\bibitem [{\citenamefont {{Foreman-Mackey}}\ \emph {et~al.}(2013)\citenamefont
  {{Foreman-Mackey}}, \citenamefont {{Hogg}}, \citenamefont {{Lang}},\ and\
  \citenamefont {{Goodman}}}]{2013PASP..125..306F}%
  \BibitemOpen
  \bibfield  {author} {\bibinfo {author} {\bibfnamefont {D.}~\bibnamefont
  {{Foreman-Mackey}}}, \bibinfo {author} {\bibfnamefont {D.~W.}\ \bibnamefont
  {{Hogg}}}, \bibinfo {author} {\bibfnamefont {D.}~\bibnamefont {{Lang}}}, \
  and\ \bibinfo {author} {\bibfnamefont {J.}~\bibnamefont {{Goodman}}},\ }\href
  {\doibase 10.1086/670067} {\bibfield  {journal} {\bibinfo  {journal} {\pasp}\
  }\textbf {\bibinfo {volume} {125}},\ \bibinfo {pages} {306} (\bibinfo {year}
  {2013})},\ \Eprint {http://arxiv.org/abs/1202.3665} {arXiv:1202.3665
  [astro-ph.IM]} \BibitemShut {NoStop}%
\bibitem [{\citenamefont {{Fong}}\ \emph {et~al.}(2014)\citenamefont {{Fong}}
  \emph {et~al.}}]{2014ApJ...780..118F}%
  \BibitemOpen
  \bibfield  {author} {\bibinfo {author} {\bibfnamefont {W.}~\bibnamefont
  {{Fong}}} \emph {et~al.},\ }\href {\doibase 10.1088/0004-637X/780/2/118}
  {\bibfield  {journal} {\bibinfo  {journal} {\apj}\ }\textbf {\bibinfo
  {volume} {780}},\ \bibinfo {eid} {118} (\bibinfo {year} {2014})},\ \Eprint
  {http://arxiv.org/abs/1309.7479} {arXiv:1309.7479 [astro-ph.HE]} \BibitemShut
  {NoStop}%
\bibitem [{\citenamefont {{{\"O}zel}}\ \emph {et~al.}(2012)\citenamefont
  {{{\"O}zel}}, \citenamefont {{Psaltis}}, \citenamefont {{Narayan}},\ and\
  \citenamefont {{Santos Villarreal}}}]{2012ApJ...757...55O}%
  \BibitemOpen
  \bibfield  {author} {\bibinfo {author} {\bibfnamefont {F.}~\bibnamefont
  {{{\"O}zel}}}, \bibinfo {author} {\bibfnamefont {D.}~\bibnamefont
  {{Psaltis}}}, \bibinfo {author} {\bibfnamefont {R.}~\bibnamefont
  {{Narayan}}}, \ and\ \bibinfo {author} {\bibfnamefont {A.}~\bibnamefont
  {{Santos Villarreal}}},\ }\href {\doibase 10.1088/0004-637X/757/1/55}
  {\bibfield  {journal} {\bibinfo  {journal} {\apj}\ }\textbf {\bibinfo
  {volume} {757}},\ \bibinfo {eid} {55} (\bibinfo {year} {2012})},\ \Eprint
  {http://arxiv.org/abs/1201.1006} {arXiv:1201.1006 [astro-ph.HE]} \BibitemShut
  {NoStop}%
\bibitem [{\citenamefont {{Abadie}}\ \emph
  {et~al.}(2012{\natexlab{b}})\citenamefont {{Abadie}} \emph
  {et~al.}}]{2012PhRvD..85h2002A}%
  \BibitemOpen
  \bibfield  {author} {\bibinfo {author} {\bibfnamefont {J.}~\bibnamefont
  {{Abadie}}} \emph {et~al.},\ }\href {\doibase 10.1103/PhysRevD.85.082002}
  {\bibfield  {journal} {\bibinfo  {journal} {\prd}\ }\textbf {\bibinfo
  {volume} {85}},\ \bibinfo {eid} {082002} (\bibinfo {year}
  {2012}{\natexlab{b}})},\ \Eprint {http://arxiv.org/abs/1111.7314}
  {arXiv:1111.7314 [gr-qc]} \BibitemShut {NoStop}%
\bibitem [{\citenamefont {{Sidery}}\ \emph {et~al.}(2014)\citenamefont
  {{Sidery}} \emph {et~al.}}]{2014PhRvD..89h4060S}%
  \BibitemOpen
  \bibfield  {author} {\bibinfo {author} {\bibfnamefont {T.}~\bibnamefont
  {{Sidery}}} \emph {et~al.},\ }\href {\doibase 10.1103/PhysRevD.89.084060}
  {\bibfield  {journal} {\bibinfo  {journal} {\prd}\ }\textbf {\bibinfo
  {volume} {89}},\ \bibinfo {eid} {084060} (\bibinfo {year} {2014})},\ \Eprint
  {http://arxiv.org/abs/1312.6013} {arXiv:1312.6013 [astro-ph.IM]} \BibitemShut
  {NoStop}%
\bibitem [{\citenamefont {{Littenberg}}\ and\ \citenamefont
  {{Cornish}}(2015)}]{2015PhRvD..91h4034L}%
  \BibitemOpen
  \bibfield  {author} {\bibinfo {author} {\bibfnamefont {T.~B.}\ \bibnamefont
  {{Littenberg}}}\ and\ \bibinfo {author} {\bibfnamefont {N.~J.}\ \bibnamefont
  {{Cornish}}},\ }\href {\doibase 10.1103/PhysRevD.91.084034} {\bibfield
  {journal} {\bibinfo  {journal} {\prd}\ }\textbf {\bibinfo {volume} {91}},\
  \bibinfo {eid} {084034} (\bibinfo {year} {2015})},\ \Eprint
  {http://arxiv.org/abs/1410.3852} {arXiv:1410.3852 [gr-qc]} \BibitemShut
  {NoStop}%
\bibitem [{\citenamefont {{Amaro-Seoane}}\ \emph {et~al.}(2013)\citenamefont
  {{Amaro-Seoane}} \emph {et~al.}}]{2013GWN.....6....4A}%
  \BibitemOpen
  \bibfield  {author} {\bibinfo {author} {\bibfnamefont {P.}~\bibnamefont
  {{Amaro-Seoane}}} \emph {et~al.},\ }\href@noop {} {\bibfield  {journal}
  {\bibinfo  {journal} {GW Notes}\ }\textbf {\bibinfo {volume} {6}},\ \bibinfo
  {pages} {4} (\bibinfo {year} {2013})},\ \Eprint
  {http://arxiv.org/abs/1201.3621} {arXiv:1201.3621 [astro-ph.CO]} \BibitemShut
  {NoStop}%
\bibitem [{\citenamefont {{Stroeer}}\ and\ \citenamefont
  {{Vecchio}}(2006)}]{2006CQGra..23S.809S}%
  \BibitemOpen
  \bibfield  {author} {\bibinfo {author} {\bibfnamefont {A.}~\bibnamefont
  {{Stroeer}}}\ and\ \bibinfo {author} {\bibfnamefont {A.}~\bibnamefont
  {{Vecchio}}},\ }\href {\doibase 10.1088/0264-9381/23/19/S19} {\bibfield
  {journal} {\bibinfo  {journal} {Class. Quantum Grav.}\ }\textbf {\bibinfo
  {volume} {23}},\ \bibinfo {pages} {S809} (\bibinfo {year} {2006})},\ \Eprint
  {http://arxiv.org/abs/astro-ph/0605227} {astro-ph/0605227} \BibitemShut
  {NoStop}%
\bibitem [{\citenamefont {Brown}\ \emph {et~al.}(2011)\citenamefont {Brown},
  \citenamefont {Kilic}, \citenamefont {Hermes}, \citenamefont {Prieto},
  \citenamefont {Kenyon},\ and\ \citenamefont {Winget}}]{2041-8205-737-1-L23}%
  \BibitemOpen
  \bibfield  {author} {\bibinfo {author} {\bibfnamefont {W.~R.}\ \bibnamefont
  {Brown}}, \bibinfo {author} {\bibfnamefont {M.}~\bibnamefont {Kilic}},
  \bibinfo {author} {\bibfnamefont {J.~J.}\ \bibnamefont {Hermes}}, \bibinfo
  {author} {\bibfnamefont {C.~A.}\ \bibnamefont {Prieto}}, \bibinfo {author}
  {\bibfnamefont {S.~J.}\ \bibnamefont {Kenyon}}, \ and\ \bibinfo {author}
  {\bibfnamefont {D.~E.}\ \bibnamefont {Winget}},\ }\href
  {http://stacks.iop.org/2041-8205/737/i=1/a=L23} {\bibfield  {journal}
  {\bibinfo  {journal} {Astrophys. J. Lett.}\ }\textbf {\bibinfo {volume}
  {737}},\ \bibinfo {pages} {L23} (\bibinfo {year} {2011})}\BibitemShut
  {NoStop}%
\bibitem [{\citenamefont {{Hermes}}\ \emph {et~al.}(2012)\citenamefont
  {{Hermes}} \emph {et~al.}}]{2012ApJ...757L..21H}%
  \BibitemOpen
  \bibfield  {author} {\bibinfo {author} {\bibfnamefont {J.~J.}\ \bibnamefont
  {{Hermes}}} \emph {et~al.},\ }\href {\doibase 10.1088/2041-8205/757/2/L21}
  {\bibfield  {journal} {\bibinfo  {journal} {Astrophys. J. Lett.}\ }\textbf
  {\bibinfo {volume} {757}},\ \bibinfo {eid} {L21} (\bibinfo {year} {2012})},\
  \Eprint {http://arxiv.org/abs/1208.5051} {arXiv:1208.5051 [astro-ph.SR]}
  \BibitemShut {NoStop}%
\bibitem [{\citenamefont {{Kilic}}\ \emph {et~al.}(2013)\citenamefont
  {{Kilic}}, \citenamefont {{Brown}},\ and\ \citenamefont
  {{Hermes}}}]{2013ASPC..467...47K}%
  \BibitemOpen
  \bibfield  {author} {\bibinfo {author} {\bibfnamefont {M.}~\bibnamefont
  {{Kilic}}}, \bibinfo {author} {\bibfnamefont {W.~R.}\ \bibnamefont
  {{Brown}}}, \ and\ \bibinfo {author} {\bibfnamefont {J.~J.}\ \bibnamefont
  {{Hermes}}},\ }in\ \href@noop {} {\emph {\bibinfo {booktitle} {9th LISA
  Symposium}}},\ \bibinfo {series} {Astronomical Society of the Pacific
  Conference Series}, Vol.\ \bibinfo {volume} {467},\ \bibinfo {editor} {edited
  by\ \bibinfo {editor} {\bibfnamefont {G.}~\bibnamefont {{Auger}}}, \bibinfo
  {editor} {\bibfnamefont {P.}~\bibnamefont {{Bin{\'e}truy}}}, \ and\ \bibinfo
  {editor} {\bibfnamefont {E.}~\bibnamefont {{Plagnol}}}}\ (\bibinfo {year}
  {2013})\ p.~\bibinfo {pages} {47}\BibitemShut {NoStop}%
\end{thebibliography}%
\end{document}